\pdfoutput=1
\RequirePackage{xcolor}
\documentclass[journal,twoside]{IEEEtran}
 
\providecommand{\journalname}{}

\usepackage{dsfont}

\usepackage{amsthm}
\usepackage{amssymb}
\usepackage{cite}
\usepackage{amsmath,amsfonts}
\allowdisplaybreaks
\usepackage[flushleft]{threeparttable}
\usepackage{textcomp}
\usepackage{graphicx}
\usepackage{xcolor}
\usepackage{algorithm}
\usepackage[capitalize]{cleveref}
\usepackage{algpseudocode}
\usepackage[T1]{fontenc}
\usepackage{times}

\algnewcommand\algorithmicforeach{\textbf{for each}}
\algdef{S}[FOR]{ForEach}[1]{\algorithmicforeach\ #1\ \algorithmicdo}

\usepackage{url}
\usepackage{bbm}
\usepackage{balance}

\usepackage[inkscape=false]{svg}
\usepackage{pgfplots}
\usepackage{pifont}
\pgfplotsset{compat=1.8}

\newtheorem*{assumption*}{Assumption}
\newtheorem{theorem}{Theorem}

\newtheorem{definition}{Definition}

\crefname{assumption}{Assumption}{Assumptions}
\crefname{corollary}{Corollary}{Corollaries}
\crefname{theorem}{Theorem}{Theorems}
\crefname{lemma}{Lemma}{Lemmas}
\crefalias{lemma}{lemma}

\usepackage{multirow}
\usepackage{overpic}
\usepackage{booktabs}
\usepackage{makecell}
\usepackage{glossaries}

\makeatletter

\let\c@lemma\c@theorem
\makeatother

\newacronym{acr:mfo}{MFO}{model-free optimisation}
\newacronym{acr:bo}{BO}{Bayesian Optimization}
\newacronym{acr:CMA-ES}{CMA-ES}{Covariance Matrix Adaptation Evolution Strategy}
\newacronym{acr:ga}{GA}{Genetic Algorithm}

\newcommand{\tup}[1]{\left(#1\right)}
\newcommand{\set}[1]{\left\{ #1 \right\}}
\newacronym{acr:pt}{PT}{public transport}
\newacronym{acr:tx}{TX}{taxi}

\newacronym{acr:mso}{MSOs}{mobility service operators}

\newcommand{\modeset}{\mathcal{M}}
\newcommand{\modeindex}{m}
\newcommand{\edgeindex}{e} 
\newcommand{\graph}{\Gamma}  
\newcommand{\nodes}{\mathcal{V}}  
\newcommand{\edges}{\mathcal{E}} 
\newcommand{\labelsmap}{\ell}
\newcommand{\labels}{\mathcal{S}}
\newcommand{\labelsingle}{s}
\newcommand{\usergroupset}{\mathcal{N}}
\newcommand{\usergroupindex}{n}
\newcommand{\vot}{\gamma}
\newcommand{\paths}{p}
\newcommand{\pathset}{\mathcal{P}}
\newcommand{\requestset}{\mathcal{R}}
\newcommand{\request}{r}
\newcommand{\requestnumber}{q_{\usergroupindex}}

\newcommand{\useraction}{\mathbf{y}}

\newcommand{\flowset}{\mathcal{Y}}
\newcommand{\flow}{\mathbf{y}}
\newcommand{\flowe}{y}
\newcommand{\flowg}{\mathbf{y_g}}
\newcommand{\flowgstar}{\mathbf{y_g^*}}
\newcommand{\flowpath}{\mathbf{f}}

\newcommand{\dispersion}{\theta}
\newcommand{\operatorset}{\mathcal{K}}
\newcommand{\operatorindex}{k}

\newcommand{\operatoractionset}{\mathcal{X}}
\newcommand{\operatoraction}{\mathbf{x}}
\newcommand{\operatorobj}{F}
\newcommand{\operatorobjf}{F}
\newcommand{\operatorgradient}{\mathbf{G}}
\newcommand{\ptfre}{q}

\newcommand{\govactionset}{\mathcal{Z}}
\newcommand{\govaction}{\mathbf{z}}
\newcommand{\govobj}{J}

\newcommand{\nonnegativenumbers}{\mathbb{R}_{\geq 0}}
\newcommand{\realnumbers}{\mathbb{R}}

\newcommand{\taxi}{\text{TX}} 
\newcommand{\pt}{\text{PT}} 
\newcommand{\costEnergy}{\phi^1} %
\newcommand{\costLabour}{\phi^2} %
\newcommand{\costVehicle}{\phi^3} %
\newcommand{\unitEmission}{\mu^1} %

\newcommand{\tax}{\tau} %
\newcommand{\license}{\lambda}       %
\newcommand{\subsidy}{\sigma} %

\newcommand{\waittimetaxi}{t^{\text{w}}_{\text{TX}}(w)} %

\newcommand{\proj}{\operatorname{Proj}}

\def\BibTeX{{\rm B\kern-.05em{\sc i\kern-.025em b}\kern-.08em
    T\kern-.1667em\lower.7ex\hbox{E}\kern-.125emX}}
\begin{document}
\title{Hierarchical Interactions in Mobility Networks: \\ A Model-Free Feedback Approach}

\author{Mingjia He$^{1,4}$, Zhiyu He$^{2}$, Jan Ghadamian$^{3}$, Florian D\"{o}rfler$^{2}$, Emilio Frazzoli$^{1}$, and Gioele Zardini$^{4}$
\thanks{This work was supported by the ETH Z\"{u}rich Mobility Initiative (MI-03-22), the ETH Z\"{u}rich Foundation (2022-HS-213), the Max Planck ETH Center for Learning Systems, and SNSF via NCCR Automation (51NF40\_225155).}
\thanks{$^{1}$Institute for Dynamic Systems and Control, ETH Z\"{u}rich, 8092 Z\"{u}rich, Switzerland
(e-mail: \text{\{minghe, efrazzoli\}@ethz.ch}).}
\thanks{$^{2}$Automatic Control Laboratory, ETH Z\"{u}rich, 8092 Z\"{u}rich, Switzerland
(e-mail: \text{\{zhiyhe, doerfler\}@control.ee.ethz.ch}).}
\thanks{$^{3}$Institute of Construction and Infrastructure Management, ETH Z\"{u}rich, 8092 Z\"{u}rich, Switzerland
(e-mail: \text{ghjan@ethz.ch}).}
\thanks{$^{4}$Laboratory for Information and Decision Systems, Massachusetts Institute of Technology, Cambridge, MA, USA
(e-mail: \text{gzardini@mit.edu}).}
}

\maketitle

\begin{abstract}
Urban mobility systems are large-scale networked systems whose performance emerges from hierarchically coupled decisions of public authorities, competing mobility operators, and travelers. 
Existing approaches to policy design either collapse one of these levels or assume that the authority can model, and differentiate through, the equilibrium responses of operators and travelers.
In this paper, we model policy design as a tri-level game on a multimodal network, in which a municipality sets taxes, licenses, and integration subsidies, a public transport and a taxi operator compete on prices and service supply, and heterogeneous travelers reach a stochastic user equilibrium.
We prove that the traveler equilibrium is unique and continuously differentiable in the operator strategies, that the variational inequality characterizing operator equilibria admits a solution for every policy, and that the resulting tri-level game admits a Stackelberg equilibrium.
Building on this structure, we design a model-free hierarchical controller in which the municipality and the operators update their decisions from zeroth-order evaluations of observed equilibrium outcomes.
On the Zurich multimodal network, operator equilibria computed from different initializations differ by a median of 0.3\% in the municipal objective.
On a grid of 3,125 policies, only those that subsidize transfers between public transport and taxis benefit both operators, and 24 of them also lower travel costs without degrading the municipal objective.
Finally, the controller matches the best policies of three black-box optimizers and terminates at the most consistent policies across runs.

\end{abstract}


\section{Introduction}
\color{black}
Urban mobility systems are large-scale socio-technical systems whose performance emerges from the interactions of self-interested stakeholders. Travelers choose routes and modes according to private preferences, mobility operators set prices, capacities, and frequencies to pursue their own objectives, and public authorities are increasingly called upon to steer the resulting system toward societal goals, such as the European Union's target of a 90\% reduction in transport-related greenhouse gas emissions by 2050~\cite{eu2020mobilitystrategy}.
Authorities, however, cannot dictate behavior. They can only reshape incentives through taxes, subsidies, and licensing constraints, whose effect depends on how operators and travelers respond.
Designing such instruments is difficult for at least three reasons.
First, the interactions are both \emph{horizontal} and \emph{vertical}.
Horizontally, heterogeneous travelers reach a multimodal user equilibrium under congestion, while operators compete in prices, capacities, and frequencies.
Vertically, municipal policy shapes the incentives under which both make their choices.
Second, the resulting equilibrium maps are nested and implicit.
Operator payoffs depend on the traveler equilibrium, the societal objective depends on both, and neither map is available in closed form.
Moreover, the operator game is nonconvex, so classical existence results do not apply and equilibria need not be unique.
Third, the authority rarely knows the operators' costs or the travelers' preferences, and observes only aggregate outcomes, such as ridership, revenues, and emissions, after the system has re-equilibrated.
Taken together, these obstacles call for design methods that treat the nested game as a plant, rely only on its observed equilibrium responses, and rest on structural properties that make the design problem well posed.
This raises the central question of this paper: \emph{How can a public authority iteratively optimize its policy, using only observed equilibrium responses, to steer a hierarchical strategic mobility system toward societal objectives?}

\subsection{Related Work}
Our work lies at the intersection of transportation research and the control of strategic multi-agent systems.

\subsubsection*{Multimodal mobility systems}
Multimodal mobility systems couple three strategic levels, namely municipal policy, operator interaction, and traveler routing under congestion, but existing work mostly models only one or two of them.
At the traveler level, congestion games are the standard lower-level model in mobility control problems such as congestion pricing and tolling~\cite{cianfanelli2023optimal, grontas2023designing} and autonomous-vehicle routing~\cite{mehr2019will,kim2025strategic}.
At the operator level, game-theoretic formulations capture competition and cooperation among self-interested providers~\cite{zardini2021game,zardini2023strategic,lanzetti2023interplay, maljkovic2025decentralized,li2026implementation,zambrano2025user}.
Representative settings include mobility-on-demand operators competing with public transit, where fleet-wide pricing can cannibalize transit ridership~\cite{lanzetti2023interplay}, competitive vehicle sharing and fleet routing~\cite{zhang2023ride, sheldon2025competitive}, and cooperative benefit allocation in multi-regional network design~\cite{he2025co}.
These works capture horizontal interactions but hold the regulatory environment fixed, although taxes, fleet-size licenses, and integration subsidies can reshape the strategic interaction itself~\cite{mo2021competition,maljkovic2023hierarchical}.

A recent line of work places a public authority above the operators, yielding a tri-level problem~\cite{gu2019tri, xi2024single, dandl2021regulating, yao2026maas, maljkovic2023hierarchical}.
Gu et al.~\cite{gu2019tri} retain competition among profit-maximizing road operators below a public agency in a static, deterministic, single-mode network, and Xi et al.~\cite{xi2024single} regulate a two-sided Mobility-as-a-Service market as a single-leader multi-follower game.
Both compute exact solutions through a mathematical program with equilibrium constraints, which requires explicit models of all followers.
Dandl et al.~\cite{dandl2021regulating} couple a welfare-maximizing regulator, a profit-maximizing provider, and an agent-based traveler simulator, using Bayesian optimization at the upper two levels, but their operator level is monopolistic.
Yao et al.~\cite{yao2026maas} collapse a multi-leader–multi-follower Mobility-as-a-Service game into a single-level variational inequality, in which the top-level agent has full model knowledge and a differentiable follower response and tunes a single wholesale price.

Our approach differs from these works in two key ways.
First, it retains strategic interaction at every level, with competing operators coupled through intermodal transfers above a congested traveler equilibrium that is neither collapsed into a single level nor available in closed form.
Second, the municipality needs no model of the lower levels and updates its policy from observed equilibrium outcomes only, while the properties established in \cref{sec:theoretical_properties} guarantee that the design problem is well posed.

\subsubsection*{Hierarchical and feedback-based control of strategic multi-agent systems} 
Settings in which a leader acts before a population of followers are commonly modeled as Stackelberg or bilevel problems~\cite{bacsar1998dynamic}.
When the followers play a non-cooperative game, the leader faces a nonconvex and nonsmooth mathematical program with equilibrium constraints, which has motivated specialized solution methods~\cite{grontas2024big, maljkovic2025decentralized}.
A parallel literature studies equilibrium seeking in network games, with convergence guarantees under mild assumptions on utilities and information~\cite{belgioioso2020distributed,yi2019operator}.

Most closely related is \emph{feedback optimization}, where a controller optimizes the steady state of a dynamical system from output measurements, without an explicit model of the input-output map~\cite{simonetto2020time,hauswirth2024optimization}, including its interconnection with equilibrium-seeking dynamics in games~\cite{belgioioso2025online}.
We adopt the same framework, but our plant is a nested game whose equilibrium maps are unavailable in closed form and may be set-valued or non-differentiable.
Existing approaches handle such maps by requiring the followers' exact cost functions~\cite{grontas2024big}, limiting lower-level action spaces~\cite{bai2021sample}, or considering a single follower rather than competing ones~\cite{kwon2023fully}.
By contrast, we use zeroth-order feedback updates at both the municipal and the operator level, which require no explicit equilibrium model.

\subsection{Contributions and Organization}
In this article, we cast urban mobility policy design as the feedback interconnection between a municipality, acting as a controller, and a nested game among operators and travelers, acting as the plant.
Our contributions are fourfold.
First, we formulate a tri-level game on a multimodal network, in which a municipality sets revenue taxes, a taxi-fleet license, and integration subsidies, a \gls{acr:pt} and a \gls{acr:tx} operator compete in prices and service supply, and heterogeneous travelers reach a stochastic user equilibrium (\cref{sec:mobility_game}).
Second, we prove that the traveler equilibrium is unique, Lipschitz continuous, and continuously differentiable in the operator strategies, that the variational inequality characterizing operator equilibria admits a solution for every policy, and that the tri-level game admits a Stackelberg equilibrium, so that structural properties propagate upward through the hierarchy (\cref{sec:theoretical_properties}).
Third, we design a model-free hierarchical controller in which the municipality and the operators update their decisions with two-point zeroth-order gradient estimates built from observed equilibrium outcomes only (\cref{sec:algorithm}).
Finally, on the multimodal network of Zurich (\cref{sec:numerical}), we show that the computed operator equilibria are reproducible across initializations, that integration subsidies are necessary for policies that benefit both operators, and that the controller attains the best grid policy while terminating at the most consistent policies among the tested black-box optimizers. \Cref{sec:conclusion} concludes the article.

A preliminary version of this work appeared in~\cite{he2025hierarchical}. This article extends it in four ways.
First, we replace the deterministic traveler equilibrium with a stochastic user equilibrium and model how the \gls{acr:pt} frequency and the \gls{acr:tx} fleet size determine waiting times.
Second, we establish the structural properties of \cref{sec:theoretical_properties}.
Third, we compute operator equilibria online with a certified stopping rule, rather than through a learned surrogate of the equilibrium map.
Finally, we extend the case study with a reliability analysis of the operator equilibria, a grid analysis of policy-induced equilibria, and a tuned benchmark against three black-box optimizers.

\section{Formulation of the Tri-Level Mobility Game}
\label{sec:mobility_game}
In this section, we formalize the tri-level mobility game, from its overview and the stakeholders' decisions (\cref{sec:overview,sec:decision_variables}) to the decision problems and equilibrium notions at the low, middle, and top level (\cref{sec:optimization_problems,sec:operator_decisions,sec:top_level_game}).

\subsection{Overview of the Mobility Game}
\label{sec:overview}
\begin{figure*}[!t]
    \centering
    \includegraphics[width=\linewidth, trim={0 0.2cm 0 0cm}, clip]{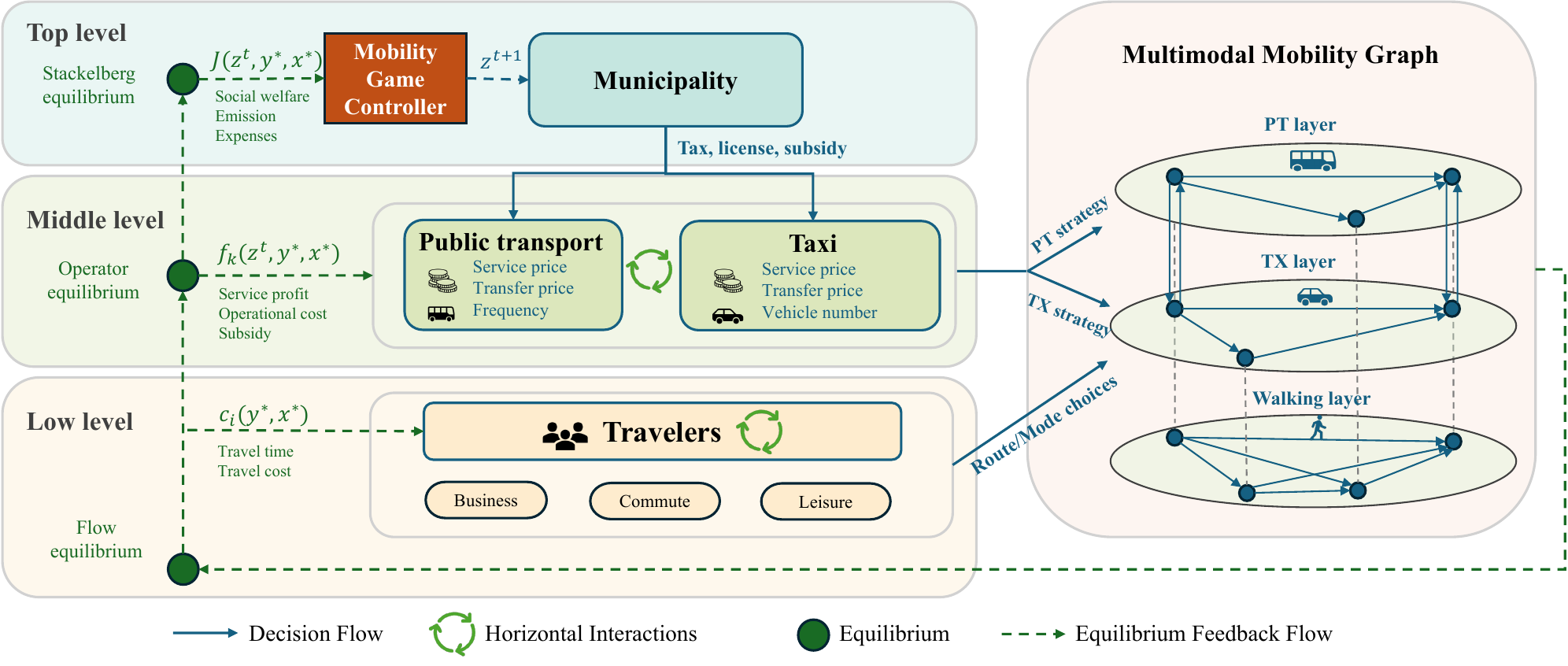}
    \caption{Hierarchical structure of the tri-level mobility game. The municipality (top level) selects the policy $\govaction$, which shapes the competition between the \gls{acr:pt} and \gls{acr:tx} operators (middle level), whose equilibrium $\operatoraction^*$ sets the edge labels on which travelers reach the equilibrium $\flow^*$ (low level). The municipality, acting as a model-free \emph{controller}, uses the observed outcomes to update $\govaction$, with a calibrated city simulator as the \emph{plant}.}
    \label{fig:mobility_game_overview}
\end{figure*}

As illustrated in \cref{fig:mobility_game_overview}, all interactions take place on a multimodal graph with layers for public transport ($\graph^1$), taxi ($\graph^2$), and walking ($\graph^3$), connected by intermodal transfer edges.
At the top level, the municipality selects a policy $\govaction = (\tax, \license, \subsidy) \in \govactionset$ of taxes, the fleet-size license, and integration subsidies.
Given $\govaction$, the \gls{acr:pt} and \gls{acr:tx} operators play a non-cooperative game over prices, transfer charges, and service supply (frequency $\ptfre$ for PT, fleet size $w$ for TX), whose equilibrium $\operatoraction^*$ determines the edge labels.
Travelers with distinct values of time then choose routes and modes, reaching a flow equilibrium $\flow^*$.
The equilibrium responses $(\operatoraction^*, \flow^*)$ propagate upward as aggregate measurements of social welfare, emissions, and fiscal revenue, which the controller of \cref{sec:algorithm} uses to update the policy $\govaction^{t} \to \govaction^{t+1}$.

\subsubsection*{Multimodal Mobility Network}
The mobility system operates on a directed labeled graph $\graph = \tup{\nodes, \edges, \labelsmap}$, where $\nodes$ is the set of vertices and $\edges \subseteq \nodes \times \nodes$ is the set of directed edges.
We consider three transportation modes $\modeindex \in \modeset = \set{1, 2, 3}$, corresponding to \gls{acr:pt}, \gls{acr:tx}, and walking, respectively.
The graph is composed of mode-specific subgraphs
$\graph^\modeindex = (\nodes^\modeindex, \edges^\modeindex, \labelsmap^\modeindex)$,
with $\nodes = \bigcup_{\modeindex} \nodes^\modeindex$ and
$\edges = \Bigl(\bigcup_{\modeindex} \edges^\modeindex\Bigr)
\cup \Bigl(\bigcup_{\modeindex \neq \modeindex'} \edges^{\modeindex \to \modeindex'}\Bigr)$,
where each $\edges^{\modeindex \to \modeindex'} \subseteq \nodes^{\modeindex} \times \nodes^{\modeindex'}$
is a set of intermodal transfer edges linking co-located nodes across layers.
The edge labeling function $\labelsmap: \edges \times \operatoractionset \times \flowset \to \labels$ assigns each edge a label $\labelsingle_e = (c_e, d_e, t_e) \in \labels = \nonnegativenumbers^3$ of monetary cost, distance, and travel time.
The labels encode the strategic coupling, since operators shape them through prices, capacities, and frequencies, and travelers through congestion.
The policy $\govaction$ enters them only indirectly, through the operators' feasible sets and incentives.
Distances $d_e$ and walking-layer labels are exogenous and fixed.
\gls{acr:pt} service edges ($e \in \edges^1$) inherit travel times from fixed timetables. In contrast, road-based edges ($e \in \edges^2$) experience congestion, with travel time modeled by the Bureau of Public Roads (BPR) function~\cite{BPR}:
\begin{equation}
\label{eq:BPR}
    t_e = t_e^0 \left(1 + a (\flowe_e/V_e)^b\right),
\end{equation}
where $t_e^0 \in \mathbb{R}_{> 0}$ is the free-flow travel time, $\flowe_e \in \mathbb{R}_{\geq 0}$ is the total flow (defined in \cref{sec:decision_variables}), $V_e \in \mathbb{R}_{> 0}$ is the edge capacity, and $a = 0.15, b = 4$ are standard BPR parameters~\cite{BPR}.
The remaining labels are set by the operators' decisions, specified next.
\subsection{Stakeholders and Decisions}
\label{sec:decision_variables}

\subsubsection*{Municipality policy instruments}
The municipality controls a policy vector $\govaction = (\tax, \license, \subsidy) \in \govactionset$ with three components:
a revenue policy $\tax = (\tax_{\pt}, \tax_{\taxi})$, where $\tax_\operatorindex \in [-1, 1]$ represents operator $\operatorindex$'s fare revenue fraction collected as a tax ($\tax_\operatorindex > 0$) or paid as a subsidy ($\tax_\operatorindex < 0$);
a license variable~$\license \in [\license_{\min}, \license_{\max}]$ capping the number of operating taxi vehicles;
and integration subsidies $\subsidy = (\subsidy_{\pt}, \subsidy_{\taxi})$ denoting per-trip payments for each intermodal transfer, capped by the service base price.
The feasible set~$\mathcal{Z}$ is the box defined by these ranges, with~$\license_{\min}\geq \underline{w}$, the minimum \gls{acr:tx} fleet size, so that the license never renders the operator's feasible set empty.

\subsubsection*{Operator strategies}
Each operator selects a strategy that determines the cost and time of its service:
\begin{align}\label{eq:operator-decision}
\operatoraction_{\pt} &= \bigl(\ptfre,\; p_{\pt}^{\text{base}},\; p_{\pt}^{\text{trans}}\bigr) \in \nonnegativenumbers^3, \nonumber \\
\operatoraction_{\taxi} &= \bigl(w,\; p_{\taxi}^{\text{base}},\; p_{\taxi}^d,\; p_{\taxi}^t,\; p_{\taxi}^{\text{trans}}\bigr) \in \nonnegativenumbers^5, \nonumber
\end{align}
forming the joint strategy profile $\operatoraction = (\operatoraction_{\pt}, \operatoraction_{\taxi}) \in \operatoractionset$, where $\operatoractionset$ is the Cartesian product of the individual feasible sets $\operatoractionset_\operatorindex$, and $\operatorindex \in \{ \pt, \taxi \}$.
The first component of each vector controls service supply.
For \gls{acr:pt}, $\ptfre$ is the service frequency in vehicles per hour; assuming uniformly spaced service and random passenger arrivals, the expected waiting time is half the headway, $t^{\mathrm{w}}_\pt(\ptfre) = 30/\ptfre$ minutes~\cite{ingvardson2018passenger}.
For \gls{acr:tx}, $w$ is the number of operating vehicles. The expected passenger waiting time follows the spatial-matching law \cite{castillo2025matching}:
\begin{equation}\label{eq:tx_waiting}
    t^{\mathrm{w}}_\taxi(w) = t^{\mathrm{w}}_0 + t^{\mathrm{w}}_\mathrm{m} \sqrt{\bar w / w},
\end{equation}
where $t^{\mathrm{w}}_0$ is the irreducible pickup and boarding time and $t^{\mathrm{w}}_\mathrm{m}$ is the matching component of the wait at a reference fleet size $\bar w$.
The remaining components define pricing: $p^{\text{base}}_\operatorindex$ is the boarding fare; $p^{d}_{\taxi}$ and $p^{t}_{\taxi}$ are per-distance and per-time fare rates for \gls{acr:tx}; and $p^{\text{trans}}_\operatorindex$ is the intermodal transfer charge applied when a passenger boards from the competing service, bounded by $p_{\operatorindex}^{\text{trans}} \leq p_{\operatorindex}^{\text{base}}$.
The mappings from operator decisions to edge labels are summarized in \cref{tab:edge-label-mapping}.

\begin{table}[t]
\centering
\caption{Mapping from operator decision variables to edge labels.}
\label{tab:edge-label-mapping}
\resizebox{\columnwidth}{!}{%
\begin{tabular}{@{}clcc@{}}
\toprule
\textbf{Operator} & \textbf{Edge set} & \textbf{cost} ($c_e$) & \textbf{time} ($t_e$) \\
\midrule
\multirow{3}{*}{PT}
  & $e \in \edges^1$ (service edge)
  & exog.
  & exog. \\
  & $e \in \edges^{3 \to 1}$ (walk $\to$ PT)
  & $p_{\pt}^{\text{base}}$
  & $30/\ptfre$ \\
  & $e \in \edges^{2 \to 1}$ (TX $\to$ PT)
  & $p_{\pt}^{\text{trans}}$
  & $30/\ptfre$ \\
\midrule
\multirow{3}{*}{TX}
  & $e \in \edges^2$ (service edge)
  & $p_{\taxi}^d \, d_e + p_{\taxi}^t \, t_e$
  & Eq.~\eqref{eq:BPR} \\
  & $e \in \edges^{3 \to 2}$ (walk $\to$ TX)
  & $p_{\taxi}^{\text{base}}$
  & $\waittimetaxi$ \\
  & $e \in \edges^{1 \to 2}$ (PT $\to$ TX)
  & $p_{\taxi}^{\text{trans}}$
  & $\waittimetaxi$ \\
\bottomrule
\end{tabular}
}
\end{table}

\subsubsection*{Traveler requests and routing}
We consider three traveler groups $\usergroupset = \{\text{Commuting},~\text{Business},~\text{Leisure}\}$ with distinct values of time $\vot^\usergroupindex \in \nonnegativenumbers$, where business travelers prioritize time, leisure travelers prioritize cost, and commuters fall in between~\cite{Axhausen2006SwissVTT}.

A trip request $\request = (\usergroupindex, o, d, \requestnumber) \in \usergroupset \times \nodes^3 \times \nodes^3 \times \nonnegativenumbers \triangleq \requestset$ specifies a traveler group, origin, and destination in the walking layer, and a demand volume.
For each request, travelers choose among a finite set $\pathset_\request$ of loopless paths from $o_\request$ to $d_\request$ in $\graph$.
We measure the generalized cost of a path $\paths \in \pathset_\request$ for request $\request \in \requestset$ in time units,
\begin{equation}\label{eq:path_cost_time}
    C^\request_\paths(\flow, \operatoraction) := \sum_{e \in \paths}
    \Bigl(
    \frac{c_e(\operatoraction)}{\vot^{\usergroupindex_\request}}
    + t_e(\flowe_e, \operatoraction)
    \Bigr),
\end{equation}
where the value of time $\vot^{\usergroupindex_\request}$ of the requesting group converts monetary costs into time, and $t_e$ depends on the total edge flow $\flowe_e$ through~\eqref{eq:BPR}.

\subsection{Low-Level Stochastic Traveler Equilibrium}
\label{sec:optimization_problems}
Given the operator strategies $\operatoraction$, travelers select routes and modes according to a stochastic user equilibrium (SUE).
Since travelers face bounded rationality, incomplete information on delays, and unobserved preferences (e.g., comfort, reliability, and mode-specific biases), we adopt a logit route-choice model with i.i.d.\ perception errors, which distributes the demand of each request across its paths in proportion to the exponential of the negative generalized cost. Since these costs depend on the flows they induce through congestion, the SUE is a fixed point.
\begin{definition}[Low-Level Stochastic Traveler Equilibrium]
\label{def:low-level Equilibrium}
Given the operator strategy profile $\operatoraction \in \operatoractionset$, a path-flow vector $\flowpath$, together with the induced group-edge flows $\flowg$ and total edge flows $\flow$, is a \emph{stochastic traveler equilibrium} if
\begin{equation}\label{eq:logit_fixed_point}
    f^\request_\paths
    = q_\request
    \frac{
    \exp\bigl(-\dispersion\, C^\request_\paths(\flow, \operatoraction)\bigr)
    }{
    \sum_{\paths' \in \pathset_\request}
    \exp\bigl(-\dispersion\, C^\request_{\paths'}(\flow, \operatoraction)\bigr)
    },
    \quad \forall\, \request \in \requestset,\; \paths \in \pathset_\request,
\end{equation}
where $\dispersion > 0$ is the dispersion parameter of the logit model.
\end{definition}

Let $\flow^* := \bigl(\sum_\usergroupindex \flowe_e^\usergroupindex\bigr)_e$, $\flowgstar := (\flowe_e^\usergroupindex)_{e,\usergroupindex}$, and $\flowpath := (f_\paths^\request)_{\paths,\request}$ collect the equilibrium edge, group-edge, and path flows, which enter the operators' and the municipality's objectives. They can be computed via the Beckmann-type convex program~\cite[Chapter~8.3.5]{Boyles2022}
\begin{subequations}\label{prob:LL}
\begin{align}
\min_{\flow,\flowg,\flowpath} ~~ & \sum_{e \in \edges^2}\int_{0}^{\flowe_e} \! t_e(s, \operatoraction)\, \mathrm{d} s
\!+\! \sum_{e \in \edges} \sum_{\usergroupindex \in \usergroupset} \! \rho^\usergroupindex \kappa_e^\usergroupindex(\operatoraction) \flowe^\usergroupindex_e
\notag
\\ & \qquad
\!+\! \frac{1}{\dispersion} \sum_{\request \in \requestset} \rho^{\usergroupindex_\request} \sum_{\paths \in \pathset_\request} \! f^\request_\paths \log f^\request_\paths
\label{eq:LL_obj} \\
\text{s.t.} ~~ & \sum_{\paths \in \pathset_\request} f^\request_\paths = q_\request, \quad \forall \request \in \requestset, \label{eq:LL_flow_conservation} \\
& \sum_{\substack{\request \in \requestset \\ \usergroupindex_\request = \usergroupindex}} \sum_{\paths \in \pathset_\request} \delta_{e}^\paths f_\paths^\request = \flowe^\usergroupindex_e, \quad \forall e \in \edges,\, \forall \usergroupindex \in \usergroupset, \label{eq:LL_edge_flow_by_group} \\
& \sum_{\usergroupindex \in \usergroupset} \flowe^\usergroupindex_e = \flowe_e, \quad \forall e \in \edges, \label{eq:LL_edge_total_flow} \\
& f^\request_\paths \geq 0, \quad \forall \request \in \requestset,\, \forall \paths \in \pathset_\request, \label{eq:LL_nonnegativity}
\end{align}
\end{subequations}
where $\delta_{e}^\paths \in \{0, 1\}$ indicates whether edge $e$ belongs to path $\paths$, and $\operatoraction$ enters through the edge costs $c_e(\operatoraction)$ and times $t_e(s, \operatoraction)$.
Because the \gls{acr:tx} time fare is charged on the congested travel time, this time enters the cost of group~$\usergroupindex$ with the group-specific coefficient $1 + p_\taxi^t/\vot^\usergroupindex$.
To obtain a congestion term common to all groups, \eqref{eq:LL_obj} expresses the costs of group~$\usergroupindex$ in units rescaled by $\rho^\usergroupindex := \vot^\usergroupindex/(\vot^\usergroupindex + p_\taxi^t)$, i.e., it multiplies both the costs and the entropy term of group~$\usergroupindex$ by $\rho^\usergroupindex$.
The choice probabilities of each group remain unchanged.
$\kappa_e^\usergroupindex$ is the flow-independent cost of edge $e$ for group~$\usergroupindex$ with $p_\taxi^d d_e/\vot^\usergroupindex$ on $\edges^2$ and $c_e/\vot^\usergroupindex + t_e$ elsewhere.
The entropy term makes the traveler equilibrium single-valued and continuously differentiable in $\operatoraction$ (\cref{prop:strict_convexity_LL,lem:LL_continuity}).
The constraints enforce flow conservation, link path flows to group and total edge flows, and impose non-negativity.

We denote by $\mathrm{LNE}: \operatoractionset \rightarrow \flowset$ the low-level equilibrium map assigning to each operator strategy the induced equilibrium flows,
$
\mathrm{LNE}(\operatoraction) := \set{
    \flow^* \in \flowset
    \;\big|\;
    \flowpath \;\text{satisfies}\; \eqref{eq:logit_fixed_point}
},
$
which is single-valued by \cref{prop:strict_convexity_LL}.

\subsection{Middle-level Game of Mobility Operators}
\label{sec:operator_decisions}

Given the policy $\govaction \in \govactionset$, the operators anticipate the travelers' equilibrium response and play a non-cooperative game.
Each operator $\operatorindex \in \operatorset = \{ \pt, \taxi \}$ solves the following problem
\begin{equation}\label{eq:middle-level-operators}
\min_{\operatoraction_\operatorindex \in \operatoractionset_\operatorindex}
\quad
\operatorobj_\operatorindex \tup{
    \operatoraction_\operatorindex,
    \operatoraction_{-\operatorindex},
    \flow^*,
    \govaction
},
\end{equation}
where $\operatorobj_\operatorindex: \operatoractionset \times \flowset \times \govactionset \rightarrow \realnumbers$ is the cost function, and $\flow^* \in \mathrm{LNE}(\operatoraction)$ is the equilibrium flow from the low level.
The operator objectives comprise three components: revenue from fare collection, operating costs, and subsidies for intermodal transfers.

The revenue of each operator is the sum of fare collections across its service and access edges:
\begingroup
\small
\begin{align}
    \operatorobjf^\mathrm{rev}_\pt(\operatoraction,\flow) &=
    \sum_{e\in\edges^{3\to1}}\flowe_e p_\pt^\mathrm{base}
    + \sum_{e\in\edges^{2\to1}}\flowe_e p_\pt^\mathrm{trans}, \nonumber \\
    \operatorobjf^\mathrm{rev}_\taxi(\operatoraction,\flow) &=
    \sum_{e\in\edges^2}\flowe_e \bigl(d_e p_\taxi^d \!+\! t_e p_\taxi^t\bigr) \notag \\
    &\quad + \sum_{e\in\edges^{3\to2}}\flowe_e p_\taxi^\mathrm{base}
    + \sum_{e\in\edges^{1\to2}}\flowe_e p_\taxi^\mathrm{trans}. \nonumber
\end{align}
\endgroup
The operating costs comprise distance-related (e.g., energy), time-related (e.g., labor), and vehicle-related fixed costs:
\begin{align}
    \operatorobjf^\mathrm{c}_{\pt} (\operatoraction) &=
    \ptfre \sum_{e \in \edges^1} \bigl(d_e \, \costEnergy_{\pt} + t_e \, \costLabour_{\pt}\bigr)
    + \costVehicle_{\pt} \, \ptfre, \nonumber\\
    \operatorobjf^\mathrm{c}_{\taxi} (\operatoraction, \flow) &=
    \sum_{e \in \edges^2} \flowe_e \bigl(d_e \, \costEnergy_{\taxi} + t_e \, \costLabour_{\taxi}\bigr)
    + \costVehicle_{\taxi} \, w,\nonumber
\end{align}
where $\costEnergy_\operatorindex$, $\costLabour_\operatorindex$, and $\costVehicle_\operatorindex \geq 0$ are the per-distance, per-time, and per-vehicle costs for operator $\operatorindex \in \operatorset$.
The cost structures reflect the asymmetry between the two modes. \gls{acr:pt} costs scale with service frequency $\ptfre$, since vehicles run on schedule regardless of ridership, whereas \gls{acr:tx} variable costs scale with the realized flow $\flow$.
The integration subsidies received from the municipality are proportional to intermodal transfer flows:
\begin{equation}
    \operatorobjf^\mathrm{s}_\pt (\flow) = \subsidy_\pt \sum_{e \in \edges^{2 \to 1}} \flowe_e, \nonumber
    \quad
    \operatorobjf^\mathrm{s}_\taxi (\flow) = \subsidy_\taxi \sum_{e \in \edges^{1 \to 2}} \flowe_e. \nonumber
\end{equation}

Combining the above components, the \gls{acr:pt} operator solves
\begin{subequations}\label{eq:ML_public_prob}
\begin{align}
    \min_{\operatoraction_\pt} \quad & \operatorobj_\pt = (\tax_\pt - 1)\, \operatorobjf^\mathrm{rev}_\pt + \operatorobjf^\mathrm{c}_\pt - \operatorobjf^\mathrm{s}_\pt + \frac{\xi_\pt}{2} \|\operatoraction_\pt\|^2 \label{eq:operator_obj_pt} \\
    \text{s.t.} \quad & \flow \in \mathrm{LNE}(\operatoraction), \\
    & \underline\operatoraction_{\pt} \leq \operatoraction_\pt \leq \overline{\operatoraction}_{\pt}. \label{eq:operator_bound_pt}
\end{align}
\end{subequations}
Similarly, the \gls{acr:tx} operator solves
\begin{subequations}\label{eq:ML_taxi_prob}
\begin{align}
    \min_{\operatoraction_\taxi} \quad & \operatorobj_\taxi = (\tax_\taxi - 1)\, \operatorobjf^\mathrm{rev}_\taxi + \operatorobjf^\mathrm{c}_\taxi - \operatorobjf^\mathrm{s}_\taxi + \frac{\xi_\taxi}{2} \|\operatoraction_\taxi\|^2 \label{eq:operator_obj_tx} \\
    \text{s.t.} \quad & \flow \in \mathrm{LNE}(\operatoraction),
    \\
    & \underline{\operatoraction}_{\taxi} \leq \operatoraction_\taxi \leq  \overline{\operatoraction}_{\taxi},
    \label{eq:operator_bound_tx}
    \\
    &  w \leq \license. \label{eq:ML_license}
\end{align}
\end{subequations}
In~\eqref{eq:operator_obj_pt} and~\eqref{eq:operator_obj_tx}, $(\tax_\operatorindex - 1) \operatorobjf^\mathrm{rev}_\operatorindex$ is the negative after-tax revenue.
The regularization with~$\xi_\operatorindex\geq 0$ can penalize extreme strategies. Our results hold for any~$\xi_\operatorindex\geq 0$, and the case study uses~$\xi_\operatorindex=0$.
The bounds $\underline\operatoraction$ and $\overline{\operatoraction}$ limit prices and supply, and~\eqref{eq:ML_license} links the \gls{acr:tx} fleet size to the license.
We denote by $\operatoractionset_\pt$ and $\operatoractionset_\taxi(\govaction)$ the resulting feasible sets, where the latter depends on $\license$, and by $\operatoractionset(\govaction) := \operatoractionset_\pt \times \operatoractionset_\taxi(\govaction)$ the joint feasible set.

\begin{definition}[Middle-Level Equilibrium]
\label{def:middle_level_equilibrium}
Given $\govaction \in \govactionset$,
a strategy profile $\operatoraction^*$ is an  equilibrium of the middle-level game if,
for every operator $\operatorindex \in \operatorset$,
\begin{equation}
\operatorobj_\operatorindex \big(
\operatoraction_\operatorindex^*,
\operatoraction_{-\operatorindex}^*,
\useraction^*,
\govaction
\big)
\!\leq\!
\operatorobj_\operatorindex \big(
\tilde{\operatoraction}_\operatorindex,
\operatoraction_{-\operatorindex}^*,
\tilde{\useraction}^*,
\govaction
\big),
\quad
\forall\, \tilde{\operatoraction}_\operatorindex \in \operatoractionset_\operatorindex(\govaction) \setminus \{ \operatoraction_\operatorindex^* \},
\label{eq:mne}
\end{equation}
where
$\useraction^*\in \mathrm{LNE}(\operatoraction^*)$
and
$\tilde{\useraction}^* \in \mathrm{LNE}
\tup{
    (\tilde{\operatoraction}_\operatorindex, \operatoraction_{-\operatorindex}^*)
}$.
\end{definition}

We use
$\mathrm{MNE}(\govaction)
:=
\set{
    \operatoraction^* \in \operatoractionset
    \;\big|\;
    \operatoraction^*
    \text{ satisfies \eqref{eq:mne}}
}$
to denote the set of middle-level equilibria given the top-level policy $\govaction$.

\subsection{Top-Level Optimization of the Municipality}
\label{sec:top_level_game}

The municipality chooses a policy vector $\govaction \in \govactionset$ to minimize a societal objective evaluated at the middle- and low-level equilibria induced by its policy:
\begin{subequations}\label{eq:top_level_municipality}
\begin{align}
\min_{\govaction, \operatoraction, \flow}
\quad &
\govobj \tup{
    \govaction,\operatoraction,\flow
} \label{eq:top_level_objective} \\
\text{s.t. } \quad & \govaction \in \govactionset,~
\operatoraction \in \mathrm{MNE}(\govaction),~\flow \in \mathrm{LNE}(\operatoraction),
\end{align}
\end{subequations}
The objective function \eqref{eq:top_level_objective} is a weighted combination of social cost, emissions, and fiscal revenue:
\begin{equation}\label{eq:municipality_prob}
    \govobj(\govaction, \operatoraction, \flow) = \omega_1 \, \govobj^\mathrm{tc}
    + \omega_2 \, \govobj^\mathrm{em}
    - \omega_3 \, \govobj^\mathrm{rev},
\end{equation}
where $\omega_1, \omega_2, \omega_3 \geq 0$ are the relative importance of each metric, reflecting the strategic priorities of the municipality.
\begin{subequations}\label{eq:municipality_objectives}
\begin{align}
    \govobj^\mathrm{tc}(\govaction, \flow) &= \sum_{\usergroupindex \in \usergroupset} \sum_{e \in \edges} \flowe_e^\usergroupindex \bigl(c_e + t_e \, \vot^\usergroupindex\bigr), \label{eq:TL_municipality_sw} \\
    \govobj^\mathrm{em}(\govaction, \flow) &= \unitEmission \sum_{e \in \edges^{2}} \flowe_e \, d_e, \label{eq:TL_emission_taxi} \\
    \govobj^\mathrm{rev}(\govaction, \operatoraction) &= \sum_{\operatorindex \in \operatorset} \Bigl(\tax_\operatorindex \, \operatorobjf^\mathrm{rev}_\operatorindex - \operatorobjf^\mathrm{s}_\operatorindex\Bigr). \label{eq:TL_municipality_rev}
\end{align}
\end{subequations}
Here, $\govobj^\mathrm{tc}$ is total system travel cost (a proxy for social welfare loss), $\govobj^\mathrm{em}$ captures distance-weighted taxi emissions with unit rate $\unitEmission$, and $\govobj^\mathrm{rev}$ is net fiscal revenue.
The objective thus depends on the decisions at all three levels.

\begin{definition}[Top-Level Stackelberg Equilibrium]\label{def:equi_tri_game}
A triplet
$
(\govaction^*, \operatoraction^*, \flow^*)
\in
\govactionset
\times
\prod_{\operatorindex \in \operatorset} \operatoractionset_\operatorindex
\times
\prod_{\edgeindex \in \edges} \flowset_\edges
$
is a \emph{Stackelberg equilibrium} of the tri-level mobility game if
    i) $\flow^* \in \mathrm{LNE}(\operatoraction^*)$,
    ii) $\operatoraction^* \in \mathrm{MNE}(\govaction^*)$, and
    iii) $\govobj(\govaction^*, \operatoraction^*, \flow^*) \leq \govobj(\govaction, \operatoraction, \flow)$ for all $\govaction \in \govactionset$, $\operatoraction \in \mathrm{MNE}(\govaction)$, and $\flow \in \mathrm{LNE}(\operatoraction)$.
\end{definition}
In words, no feasible policy, together with any of the operator and traveler equilibria it induces, achieves a lower societal cost. Equivalently, $(\govaction^*, \operatoraction^*, \flow^*)$ solves~\eqref{eq:top_level_municipality}. Since~\eqref{eq:top_level_municipality} also minimizes over the induced equilibria, \cref{def:equi_tri_game} corresponds to the \emph{optimistic} Stackelberg notion, in which ties among follower equilibria are broken in favor of the leader.

\section{Properties of the Mobility Game}
\label{sec:theoretical_properties}

In this section, we establish structural properties that propagate upward through the hierarchy and make the design problem of \cref{sec:algorithm} well posed. We show that the traveler equilibrium is unique, Lipschitz continuous, and continuously differentiable in the operator strategies (\cref{sec:LL_analysis}), that the operator variational inequality admits a solution for every policy (\cref{sec:ML_analysis}), and that a Stackelberg equilibrium exists (\cref{sec:TL_analysis}).

\subsection{Properties of Low-Level Traveler Equilibrium}
\label{sec:LL_analysis}

Problem~\eqref{prob:LL} extends the classical Beckmann-type program for stochastic assignment~\cite[Chapters~5 and 8]{Boyles2022} with group-specific monetary costs.

\begin{theorem}[Existence and Uniqueness of Equilibrium Flows]
\label{prop:strict_convexity_LL}
For problem~\eqref{prob:LL}, the aggregate equilibrium group-edge flow vector~$\flowgstar = (\flowe_e^\usergroupindex)_{e,\usergroupindex}$, edge flow vector $\flow^* = \bigl(\sum_\usergroupindex \flowe_e^\usergroupindex\bigr)_e$, and path-flow vector $\flowpath = (f^\request_\paths)_{\paths,\request}$ exist and are unique.
\end{theorem}

\begin{proof}
The constraints~\eqref{eq:LL_flow_conservation}--\eqref{eq:LL_nonnegativity} are affine, so the feasible set of~\eqref{prob:LL} is convex.

Since $\flowe_e$, $\flowe_e^\usergroupindex$, and $\flowgstar$ are linear in $\flowpath$, we view~\eqref{prob:LL} as a problem in $\flowpath$. Let $\Phi_e(\flowe_e) := \int_0^{\flowe_e}t_e(s)\, \mathrm{d} s$. For \gls{acr:tx} service edges ($e \in \edges^2$), $t_e$ follows the BPR function~\eqref{eq:BPR}, so $\Phi_e$ is strictly convex. For $e \in \edges \setminus \edges^2$, $t_e$ does not depend on $\flowe_e$ and enters the linear second term through $\kappa_e^\usergroupindex$. Hence, the first term in~\eqref{eq:LL_obj}, a non-negative sum of convex functions composed with affine maps, is convex in $\flowpath$.

The second term is linear in $\flowpath$. The Hessian of the third term is diagonal with entries $\rho^{\usergroupindex_\request}/(\dispersion f^\request_\paths) \geq \underline{\rho}/(\dispersion \bar{q})$, where $\bar{q} := \max_{\request \in \requestset} q_\request$ and $\underline{\rho} := \min_{\usergroupindex \in \usergroupset} \frac{\vot^\usergroupindex}{\vot^\usergroupindex + \bar p^t_\taxi} > 0$, by~\eqref{eq:LL_flow_conservation} and~\eqref{eq:LL_nonnegativity}, so this term is $\underline{\rho}/(\dispersion \bar{q})$-strongly convex in $\flowpath$.

Therefore, the objective is $\underline{\rho}/(\dispersion \bar{q})$-strongly convex in $\flowpath$, and~\cite[Proposition~3.3]{Boyles2022} yields the existence and uniqueness of $\flowpath$, and hence of the linearly dependent flows $\flowgstar$ and $\flow^*$.
\end{proof}

We now study how the equilibrium flow depends on the operator decision $\operatoraction \in \nonnegativenumbers^{8}$.

\begin{theorem}[Lipschitz Continuity and Differentiability of Equilibrium Flow]
\label{lem:LL_continuity}
Given problem~\eqref{prob:LL}, the aggregate equilibrium flow $\flowgstar(\operatoraction)$ is $L_\flow$-Lipschitz continuous on $\operatoractionset$, i.e., $\forall \operatoraction, \operatoraction' \in \operatoractionset, \|\flowgstar(\operatoraction) - \flowgstar(\operatoraction')\| \leq L_\flow \|\operatoraction - \operatoraction'\|$,
where $L_\flow$ is a positive constant. Further, $\flowgstar(\operatoraction)$ is continuously differentiable with respect to $\operatoraction$.

\end{theorem}

\begin{proof}
The proof relies on the sensitivity analysis of monotone variational inequalities~\cite[Thm.~2F.6, 2F.7]{dontchev2009implicit}. Since $\flowgstar$ is linear in $\flowpath$, it suffices to prove both properties for $\flowpath$.

Let $\Psi(\flowpath, \operatoraction)$ denote the objective~\eqref{eq:LL_obj} written in terms of $\flowpath$ using~\eqref{eq:LL_flow_conservation}--\eqref{eq:LL_edge_total_flow}.
By the proof of \cref{prop:strict_convexity_LL}, $\Psi$ is $\underline{\rho}/(\dispersion \bar{q})$-strongly convex in $\flowpath$ over a convex feasible set, so for $\operatoraction, \operatoraction' \in \operatoractionset$ the optimal solutions satisfy
\begin{subequations}
\begin{align}
\langle \nabla_\flowpath \Psi(\flowpath(\operatoraction), \operatoraction),\; \flowpath(\operatoraction') - \flowpath(\operatoraction) \rangle &\geq 0, \label{eq:vi_action_1} \\
\langle \nabla_\flowpath \Psi(\flowpath(\operatoraction'), \operatoraction'),\; \flowpath(\operatoraction) - \flowpath(\operatoraction') \rangle &\geq 0. \label{eq:vi_action_2}
\end{align}
\end{subequations}
Moreover, the strong monotonicity of $\nabla_\flowpath \Psi(\cdot, \operatoraction)$ gives
\begin{align}\label{eq:strong_monotone_grad_Psi}
    \langle \nabla_\flowpath \Psi(\flowpath(\operatoraction), \operatoraction) &- \nabla_\flowpath \Psi(\flowpath(\operatoraction'), \operatoraction),\; \flowpath(\operatoraction) - \flowpath(\operatoraction') \rangle \notag \\
    &\geq \underline{\rho}/(\dispersion \bar{q}) \cdot \|\flowpath(\operatoraction) - \flowpath(\operatoraction')\|^2.
\end{align}

We next show that $\nabla_\flowpath \Psi(\flowpath, \operatoraction)$ is Lipschitz continuous in $\operatoraction$ at equilibrium path flows. Its entry for path $\paths$ of request $\request$ is
\begin{equation*}
\frac{\partial \Psi}{\partial f^\request_\paths}
= \!\sum_{e \in \edges^2}\! \delta_e^\paths t_e(\flowe_e, \operatoraction)
+ \rho^{\usergroupindex_\request} \!\sum_{e \in \edges}\! \delta_e^\paths \kappa_e^{\usergroupindex_\request}(\operatoraction)
+ \tfrac{\rho^{\usergroupindex_\request}}{\dispersion} (\log f^\request_\paths + 1),
\end{equation*}
By~\eqref{eq:LL_edge_flow_by_group} and~\eqref{eq:LL_edge_total_flow}, $|\partial \flowe^\usergroupindex_e / \partial f^\request_\paths| \leq 1$ and $|\partial \flowe_e/ \partial f^\request_\paths| \leq 3$ (three traveler groups).
The partial derivatives with respect to the components of $\operatoraction$ satisfy $|\partial c_e/\partial p^{\text{base/trans}}_\operatorindex| = 1$ for flat fares, $|\partial t_e/\partial \ptfre| = 30/\ptfre^2 \leq 30/\underline{\ptfre}^2$ for the \gls{acr:pt} frequency with lower bound $\underline{\ptfre}$, $|\partial c_e/\partial p_\taxi^d| = d_e \leq \bar d$ with $\bar d$ the maximum distance, $|\partial \rho^\usergroupindex/\partial p_\taxi^t| \leq 1/\vot_{\min}$, which multiplies $\kappa_e^\usergroupindex + (\log f^\request_\paths + 1)/\dispersion$, at most $\bar t := \bar\kappa + (1 + \max\{|\log \underline{f}|, |\log \bar q|\})/\dispersion$. 
Here, $\kappa_e^\usergroupindex \leq \bar\kappa := \bar c/\vot_{\min} + \bar t^{\,0}$, where $\bar c$ is the largest flow-independent edge fare and $\bar t^{\,0}$ the largest flow-independent edge time (timetable, walking, and waiting times), and $\underline{f} := \min_{\request} q_\request e^{-\dispersion \bar C}/|\pathset_\request|$ bounds the equilibrium path flows from below by~\eqref{eq:logit_fixed_point}, with $\bar C$ the largest path cost. Finally, by~\eqref{eq:tx_waiting}, $|\partial t_e/\partial w| \leq L_{w} := t^\mathrm{w}_\mathrm{m} \sqrt{\bar w} / (2 \underline{w}^{3/2})$.
Hence, the Lipschitz constant $L_\Psi^\operatoraction$ of $\nabla_\flowpath \Psi$ in $\operatoraction$ is
\begin{equation*}%
    2 \sqrt{2} |\edges| |\requestset| |\pathset_\request| \max\Bigl(\!|\usergroupset| \Bigl\{\!
    \tfrac{\bar d}{\vot_{\min}},
    \tfrac{\bar t}{\vot_{\min}},
    \tfrac{1}{\vot_{\min}}\!
    \Bigr\} + \max\Bigl\{\! 3L_{w}, \tfrac{90}{\underline{\ptfre}^2} \!\Bigr\}\Bigr),
\end{equation*}
where $\vot_{\min} := \min_{\usergroupindex \in \usergroupset} \vot^\usergroupindex$ is the minimum value of time across all traveler groups.
We sum up~\eqref{eq:vi_action_1},~\eqref{eq:vi_action_2}, and~\eqref{eq:strong_monotone_grad_Psi} and arrive at
\begin{align*}
    \underline{\rho}/(\dispersion \bar{q}) &\cdot \|\flowpath(\operatoraction) - \flowpath(\operatoraction')\|^2 \\
    &\leq \langle \nabla_\flowpath \Psi(\flowpath(\operatoraction'), \operatoraction) \!-\! \nabla_\flowpath \Psi(\flowpath(\operatoraction'), \operatoraction'),\; \flowpath(\operatoraction) \!-\! \flowpath(\operatoraction') \rangle \\
    &\leq L_\Psi^\operatoraction\,\|\operatoraction - \operatoraction'\|\,\|\flowpath(\operatoraction) - \flowpath(\operatoraction')\|,
\end{align*}
by the Cauchy--Schwarz inequality. Dividing by $\|\flowpath(\operatoraction) - \flowpath(\operatoraction')\|$ shows that $\flowpath(\operatoraction)$ is Lipschitz continuous with constant $L_\Psi^\operatoraction \dispersion \bar{q}/\underline{\rho}$, and hence $\flowgstar(\operatoraction)$ with a constant $L_\flow$ proportional to $L_\Psi^\operatoraction \dispersion \bar{q}/\rho$.

For differentiability, fix $\operatoraction$ and consider~\eqref{eq:LL_flow_conservation}, since~\eqref{eq:LL_edge_flow_by_group} and~\eqref{eq:LL_edge_total_flow} only define edge flows. Its Jacobian with respect to $\flowpath$ has rows with disjoint supports, each nonzero since every request has a feasible path, so it has full row rank and the linear independence constraint qualification holds.
Moreover, by the logit fixed point~\eqref{eq:logit_fixed_point}, every equilibrium path flow is strictly positive, so the non-negativity constraints~\eqref{eq:LL_nonnegativity} are inactive at the solution and the optimality conditions reduce to a system of equations.
Together with the strong convexity of the objective,~\cite[Thm.~1B.1]{dontchev2009implicit} and~\cite[Thm.~5.1]{fiacco1990sensitivity} imply that~$\flowpath$, and hence~$\flowgstar$, is continuously differentiable in a neighborhood of~$\operatoraction$.
\end{proof}

In words, every operator strategy profile induces a unique traveler flow pattern, which changes smoothly with prices and service supply. This makes the operator objectives continuously differentiable along the traveler response, as required at the middle level. The constant $L_\flow$ is conservative, since it grows with the numbers of edges, requests, and paths.

\subsection{Existence of Variational Inequality Solution for the Middle-Level Game}
\label{sec:ML_analysis}

The operator objectives~\eqref{eq:operator_obj_pt} and~\eqref{eq:operator_obj_tx} are nonconvex in general, since the revenues contain bilinear terms and the equilibrium flows depend nonlinearly on $\operatoraction$. Hence, the classical existence results for Nash equilibria of convex games do not apply, and we instead study the variational inequality associated with the first-order optimality conditions of~\eqref{eq:ML_public_prob} and~\eqref{eq:ML_taxi_prob}.
Every Nash equilibrium of the middle-level game solves this variational inequality~\cite[Proposition~2.1]{mertikopoulos2019learning}. Since the operator objectives are nonconvex, the converse does not hold in general, and a solution of the variational inequality is a first-order (stationary) equilibrium rather than a Nash equilibrium.
Let the pseudo-gradient mapping of \gls{acr:pt} and \gls{acr:tx} be
\begin{equation*}
    \operatorgradient(\operatoraction, \govaction) :=
    \begin{pmatrix}
    \nabla_{\operatoraction_\pt}\, \operatorobj_\pt(\operatoraction, \flow^*(\operatoraction), \govaction) \\
    \nabla_{\operatoraction_\taxi}\, \operatorobj_\taxi(\operatoraction, \flow^*(\operatoraction), \govaction)
    \end{pmatrix}.
    \label{eq:pseudo-gradient}
\end{equation*}
Here,~$\nabla_{\operatoraction_\operatorindex}$ denotes the gradient of the composite map $\operatoraction_\operatorindex \mapsto \operatorobj_\operatorindex(\operatoraction, \flow^*(\operatoraction), \govaction)$, which accounts for the traveler response and exists by \cref{lem:LL_continuity}.

For a given policy $\govaction \in \govactionset$, we consider the variational inequality
\begin{equation}\label{eq:vi_joint}
\text{Find } \operatoraction^* \!\in\! \operatoractionset(\govaction) \text{ such that } \langle \operatorgradient(\operatoraction^*, \govaction), \operatoraction' \!-\! \operatoraction^* \rangle \!\geq\! 0, ~ \forall \operatoraction' \!\in\! \operatoractionset(\govaction).
\end{equation}
The following theorem establishes that~\eqref{eq:vi_joint} admits a solution, i.e., that candidate operator equilibria exist for every policy.

\begin{theorem}[Existence of Variational Inequality Solution]\label{thm:existence_VI}
With the equilibrium flows solved from~\eqref{prob:LL}, the variational inequality~\eqref{eq:vi_joint} associated with the middle-level game admits a solution for every $\govaction \in \govactionset$.
\end{theorem}

\begin{proof}
We follow the standard argument based on Brouwer's fixed-point theorem~\cite[Theorem 2A.1]{dontchev2009implicit},~\cite[Proposition 2.2.3]{facchinei2003finite}. For any fixed $\govaction$, the linear constraints~\eqref{eq:operator_bound_pt},~\eqref{eq:operator_bound_tx}, and~\eqref{eq:ML_license} make $\operatoractionset(\govaction)$ non-empty (since $\license \geq \license_{\min} \geq \underline{w}$), convex, and compact, and $\operatorgradient(\cdot, \govaction)$ is continuous, since the operator objectives and $\flow^*(\operatoraction)$ are continuously differentiable. Brouwer's fixed-point theorem then yields a solution of~\eqref{eq:vi_joint}.
\end{proof}

In words, for every policy, there exists an operator strategy profile at which no operator can improve its objective to first order by a feasible unilateral deviation. \Cref{thm:existence_VI} does not guarantee that this profile is a Nash equilibrium, nor that it is unique. Empirically, the profiles computed by our controller from different initial strategies differ by a median of $0.3\%$ in the municipal objective (\cref{sec:numerical}).

\subsection{Existence of Top-Level Stackelberg Equilibrium}
\label{sec:TL_analysis}

By \cref{thm:existence_VI}, the solution map $\mathrm{SOL}(\operatoractionset(\govaction), \operatorgradient(\cdot, \govaction))$ of~\eqref{eq:vi_joint} is non-empty for every $\govaction \in \govactionset$. We therefore use its elements as surrogates for middle-level Nash equilibria in~\eqref{eq:top_level_municipality}, which yields
\begin{subequations}\label{eq:top_level_municipality_reformulated}
\begin{align}
\min_{\govaction, \operatoraction, \flow}
\quad &
\govobj \tup{
    \govaction,\operatoraction,\flow
} \\
\text{s.t. } \quad & \govaction \in \govactionset,~
 \operatoraction \in{}\mathrm{SOL}(\operatoractionset(\govaction), \operatorgradient(\cdot, \govaction)),~\flow \in \mathrm{LNE}(\operatoraction).
\end{align}
\end{subequations}

\begin{theorem}[Existence of Stackelberg Equilibrium]
\label{thm:existence_stackelberg}
Given the low-level and middle-level games in \Cref{sec:optimization_problems,sec:operator_decisions}, respectively, there exists a Stackelberg equilibrium $(\govaction^*, \operatoraction^*, \flow^*)$ of the overall problem~\eqref{eq:top_level_municipality_reformulated}.
\end{theorem}

\begin{proof}
The set $\govactionset$ is compact and convex, and, by the proof of \cref{thm:existence_VI}, $\operatorgradient(\cdot,\govaction)$ is continuous for every $\govaction \in \govactionset$.
Moreover, $\operatorgradient$ is jointly continuous in $(\operatoraction, \govaction)$, since the policy variables $\tax$ and $\subsidy$ enter the operator objectives linearly, and the set-valued map $\govaction \mapsto \operatoractionset(\govaction)$ is continuous with non-empty compact values, since $\license$ enters only as the upper bound $w \leq \license$ with $\license \geq \underline{w}$.
Therefore, the solution map of~\eqref{eq:vi_joint} is upper-semicontinuous and, since $\operatoractionset(\govaction)$ is non-empty and compact, closed-valued. By the closed graph theorem~\cite[Theorem~17.11]{aliprantis2006infinite}, its graph is a closed subset of $\govactionset \times \operatoractionset$. Since $\govobj$ is continuous, the extreme value theorem yields an optimal policy $\govaction^*$ for~\eqref{eq:top_level_municipality_reformulated}, which, with the corresponding $\operatoraction^* \in \mathrm{SOL}(\operatoractionset(\govaction^*), \operatorgradient(\cdot, \govaction^*))$ and $\flow^* \in \mathrm{LNE}(\operatoraction^*)$, constitutes a Stackelberg equilibrium of the overall game.
\end{proof}

\Cref{thm:existence_stackelberg} guarantees that the municipality's design problem is well posed, in the sense that an optimal policy exists. It does not provide a way to compute it, since the reduced objective is nonconvex and each of its evaluations requires solving the nested equilibrium problems. This motivates the model-free controller of the next section.

\section{Model-Free Hierarchical Controller}
\label{sec:algorithm}

Solving the hierarchical mobility game~\eqref{eq:top_level_municipality} requires handling nested equilibrium maps without closed-form expressions.
Furthermore, the resulting optimization landscape is non-smooth and non-convex, which challenges classical differentiation-based algorithms~\cite{blondel2024elements}.
We address these challenges via model-free feedback optimization~\cite{Zhiyu2024}. The key ideas are to leverage the responses of lower-level stakeholders as feedback and to construct gradient estimates from objective evaluations at structurally perturbed points. These designs facilitate iterative policy adjustments (of the top-level municipality) and equilibrium seeking (of the middle-level operators), bypassing explicit equilibrium mappings and inter-level interaction models.
Throughout, we write $J(\govaction)$ for the reduced municipal objective, i.e., the value of $\govobj(\govaction, \operatoraction, \flow)$ at the operator equilibrium $\operatoraction$ reached by the operator loop of \cref{alg:hierarchical} and at the traveler equilibrium $\flow \in \mathrm{LNE}(\operatoraction)$. Since the operator game may admit multiple equilibria (\cref{sec:ML_analysis}), $J(\govaction)$ depends on this selection, and \cref{sec:numerical} quantifies its sensitivity to the initial operator strategies.

\begin{algorithm}[t]
\caption{Model-free iterations for the hierarchical game}
\label{alg:hierarchical}
\small
\begin{algorithmic}[1]
\Require Initial policy $\govaction^0$ and operator strategies $\operatoraction^0 = (\operatoraction_\pt^0, \operatoraction_\taxi^0)$; horizons $T, S$; step sizes $\eta_\govaction, \eta_\operatoraction$; {exploration radii} $\delta_\govaction, \delta_\operatoraction$; tolerance {$\varepsilon_\mathrm{MNE}$}. \textsc{SUE} uses path set $\mathcal{P}$, dispersion $\theta$, and tolerance $\varepsilon_\mathrm{SUE}$.
\For{$t = 0, \ldots, T-1$} \Comment{municipality updates}
    \State Sample $v^t \sim \mathcal{N}(0, I)$
    \State $\govaction^\pm \leftarrow \proj_{\govactionset}(\govaction^t \pm \delta_\govaction v^t)$ \Comment{exploration policies}
    \For{$\govaction' \in \{\govaction^+, \govaction^-\}$}
        \State $\operatoraction \leftarrow \hat{\operatoraction}(\tilde{\govaction})$, \; $\tilde{\govaction} \in \arg\min_{\govaction \in \mathcal{Z}_{\mathrm{ev}}} \|\govaction - \govaction'\|$ \Comment{warm start} \label{line:warm_start}
        \State $\bar\flow \leftarrow \textsc{SUE}(\operatoraction, \govaction')$
        \For{$s = 0, \ldots, S-1$} \Comment{operator interactions} \label{line:br_start}
            \For{$\operatorindex = \taxi, \pt$} \Comment{{update against latest} $\operatoraction_{-\operatorindex}$}
                \State Sample $u_\operatorindex^s \sim \mathcal{N}(0, I)$
                \State $\flow^\pm \leftarrow \textsc{SUE}\bigl((\operatoraction_\operatorindex \pm \delta_\operatoraction u_\operatorindex^s,\, \operatoraction_{-\operatorindex}), \govaction';\, \bar\flow\bigr)$ \label{line:sue}
                \State $\operatorobj_\operatorindex^\pm \leftarrow \operatorobj_\operatorindex(\operatoraction_\operatorindex \pm \delta_\operatoraction u_\operatorindex^s,\, \operatoraction_{-\operatorindex}, \flow^\pm, \govaction')$ \label{line:operator_objective}
                \State $\operatoraction_\operatorindex \leftarrow \proj_{\operatoractionset_\operatorindex}\bigl(\operatoraction_\operatorindex - \frac{\eta_\operatoraction}{2\delta_\operatoraction}\, (\operatorobj_\operatorindex^+ - \operatorobj_\operatorindex^-)\, u_\operatorindex^s\bigr)$ \label{line:operator_update}
            \EndFor
            \State \algorithmicif\ $r(\operatoraction) \le {\varepsilon_\mathrm{MNE}}$\ \algorithmicthen\ \textbf{break} \label{line:br_end}
        \EndFor 
        \State $\widehat\govobj(\govaction') \leftarrow \govobj\bigl(\govaction', \operatoraction, \textsc{SUE}(\operatoraction, \govaction'; \bar\flow)\bigr)$ \label{line:probe_policies}
    \EndFor
    \State $\govaction^{t+1} \leftarrow \proj_{\govactionset}\bigl(\govaction^t - \frac{\eta_\govaction}{2\delta_\govaction}\, \bigl(\widehat\govobj(\govaction^+) - \widehat\govobj(\govaction^-)\bigr)\, v^t\bigr)$ \label{line:policy_update}
\EndFor
\State \Return $\govaction^T$
\end{algorithmic}
\end{algorithm}

\Cref{alg:hierarchical} summarizes the controller, which consists of three nested loops. The outer loop updates the municipality's policy, and the middle loop lets
the operators adjust their strategies until the operator game
reaches an equilibrium. The inner routine \textsc{SUE} computes the traveler equilibrium for a given middle-level strategy. 
\textsc{SUE} is a method of successive averages with logit loading, the standard way to compute a stochastic user equilibrium~\cite{liu2009method}. 
The outer and middle loops are model-free, since they use only objective values at perturbed points rather than the equilibrium maps or their derivatives.
At iteration $t$, the municipality draws $v^t \sim \mathcal{N}(0, I)$, lets the operators and travelers respond to the two exploration policies $\govaction^\pm$, and takes a projected step along the resulting two-point estimate (lines~\ref{line:probe_policies}--\ref{line:policy_update}). The operators update in the same way, in turn, each against the other's latest strategy (lines~\ref{line:br_start}--\ref{line:br_end}), and every operator evaluation calls \textsc{SUE} (line~\ref{line:sue}).

\subsubsection*{Gradient estimates}
For the top-level municipality and middle-level operators, the direct construction of gradients is challenging due to nested equilibrium mappings and limited knowledge of parameters from lower-level stakeholders. 
Instead, we use function evaluations at perturbed points to estimate gradients, thereby achieving model-free iterations. 
When the perturbed policies lie in $\govactionset$ and the inner equilibria are computed exactly, the two-point estimator used in line~\ref{line:policy_update}, i.e., $\bigl(\widehat\govobj(\govaction^+) - \widehat\govobj(\govaction^-)\bigr) v^t / (2\delta_\govaction)$, is an unbiased estimate of the gradient of the Gaussian-smoothed surrogate $J_{\delta_\govaction}(\govaction) := \mathbb{E}_{v\sim \mathcal{N}(0, I)}[J(\govaction + \delta_\govaction v)]$, whose deviation from the true objective $J(\govaction)$ is controlled by the exploration parameter $\delta_\govaction$~\cite{Zhiyu2024}. The same intuition applies to the operator level (line~\ref{line:operator_update}).

\subsubsection*{Stopping rule and warm starts}
The operator loop stops on first-order residuals of the operator game (line~\ref{line:br_end}), with $r(\operatoraction) := \max\{r_\mathrm{VI}, r_\taxi, r_\pt\}$. 
The joint $r_\mathrm{VI}(\operatoraction) := \|\operatoraction - \proj_{\operatoractionset}(\operatoraction - \widehat{\operatorgradient}(\operatoraction, \govaction))\| / \sqrt{n_\operatoraction}$ uses an estimate $\widehat{\operatorgradient}$ of the pseudo-gradient $\operatorgradient$, where $n_\operatoraction$ is the number of operator decision variables. 
The individual residual of operator $\operatorindex$ is $r_\operatorindex(\operatoraction) := \|\operatoraction_\operatorindex - \proj_{\operatoractionset_\operatorindex}(\operatoraction_\operatorindex - \widehat\nabla_{\operatoraction_\operatorindex}\operatorobj_\operatorindex)\|$. 
Both $\widehat{\operatorgradient}$ and $\widehat\nabla_{\operatoraction_\operatorindex}\operatorobj_\operatorindex$ are zeroth-order estimates, each of which requires solving the traveler equilibrium.
A profile that passes all three tests approximately solves the variational inequality~\eqref{eq:vi_joint}, and we refer to it as a \emph{certified operator equilibrium}.
Two warm starts reduce the cost per evaluation: the operator loop at a new policy starts from the equilibrium $\hat{\operatoraction}(\tilde{\govaction})$ of the nearest evaluated policy $\tilde{\govaction} \in \mathcal{Z}_{\mathrm{ev}}$ (line~\ref{line:warm_start}), or from $\operatoraction^0$ if none exists, and every call to \textsc{SUE} starts from the previous flow.

\section{Case Study on the Zurich Multimodal Network}\label{sec:numerical}
\begin{figure}
    \centering
    \includegraphics[width=0.8\linewidth]{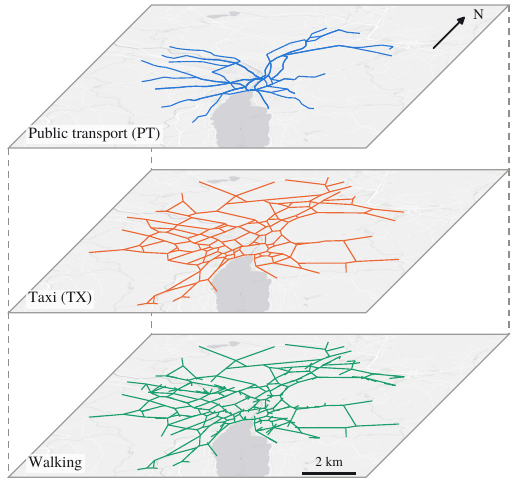}
    \caption{Multimodal network of Zurich.}
    \label{fig:map}
\end{figure}

We evaluate the presented formulation and the model-free hierarchical controller on the mobility network of Zurich, Switzerland, as shown in~\cref{fig:map}.
This multimodal network is built from OpenStreetMap~\cite{openstreetmap}, including road, walking, and tram layers.
The resulting multimodal graph has 891 vertices and 4,244 edges.
Travel demand corresponds to a one-hour slice from a full-day simulation calibrated with Swiss Federal Statistical Office population data~\cite{sbfsPortal}, yielding 1,928 distinct trips.
All stakeholder parameters, including values of time, operator cost, price ranges, and the feasible policy set, are calibrated to real-world conditions. 
We report these parameters, including the logit dispersion $\dispersion$ and the operator regularization $\xi$, in \cref{table:model_parameters} in Appendix~\ref{sec:appendix}.

Our goals are threefold. First, we assess whether the operator loop of \cref{alg:hierarchical} reliably reaches operator equilibria, and whether these equilibria depend on the initial strategies. Second, we examine how municipal policy reshapes the operator equilibria, and which instruments make both operators better off. Third, we evaluate the full controller on the tri-level problem against three black-box optimizers.
All experiments use the unregularized game ($\xi = 0$). We regard a strategy profile as an equilibrium if its residual $r(\operatoraction)$ (\cref{sec:algorithm}) is at most $\varepsilon_\mathrm{MNE} = 0.03$. All runs are implemented in Python 3.10 and use a single core of an AMD EPYC 7763 node on the ETH Zurich Euler cluster.

\subsubsection*{Convergence and Reproducibility of Operator Equilibria}\label{sec:reliability}
To evaluate the convergence of the operator loop, we sample $100$ grid policies via Latin hypercube sampling~\cite{loh1996latin}, ensuring an even distribution of both the individual policy levels (each appearing $20$ times) and their combinations. We then evaluate each policy from eight random initial operator strategies and compare the resulting certified equilibria.

\begin{figure}[tb]
    \centering
    \includegraphics[width=0.85\linewidth]{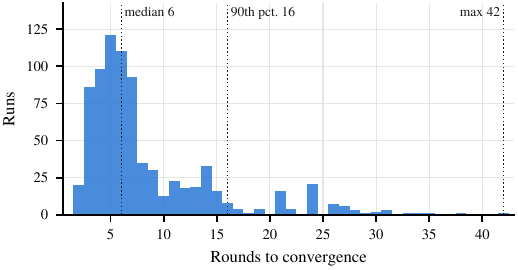}
    \caption{Number of rounds until the operator loop meets the stopping rule, over $800$ runs ($100$ policies, eight initial operator strategies each). Dashed lines mark the median, the 90th percentile, and the maximum.}
    \label{fig:convergence_ml}
\end{figure}

Across $800$ total runs, the median run converges in $6$ rounds, $90\%$ within $16$ rounds and all within $42$ rounds (\cref{fig:convergence_ml}).
The computed equilibria are also robust to the initial strategies. Across the eight initializations for a given policy, the median relative spread is $0.3\%$ for the municipality objective, and $1.2\%$ and $1.7\%$ for the \gls{acr:tx} and \gls{acr:pt} profits, respectively. \Cref{fig:br_unique} in Appendix~\ref{app:www} shows the operator iterations from eight initial strategies under three selected policies. Overall, these results indicate that, although \cref{thm:existence_VI} does not guarantee uniqueness, the equilibrium selected by the operator loop has little influence on the municipal objective in this case study.

\subsubsection*{Impact of Municipal Policy on Middle-Level Equilibria}\label{subsec:policy_mid_level}
\begin{figure}[tb]
    \centering
    \includegraphics[width=0.85\linewidth]{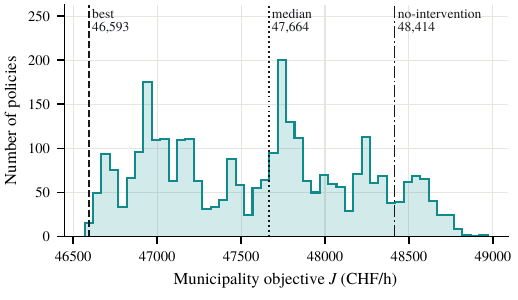}
    \caption{Distribution of the municipality objective over the $3{,}125$ policies of the $5^5$ grid. Dashed lines mark the best and the median grid policy and the no-intervention policy.}
    \label{fig:pareto}
\end{figure}

We examine how municipal policy reshapes the middle-level interaction by evaluating the policy space $\govactionset$ on a grid of five levels per policy variable over the ranges in \cref{table:model_parameters}, i.e., $5^5 = 3{,}125$ policies. For each policy, the operators iterate as specified in \cref{alg:hierarchical} until the stopping rule certifies an equilibrium, which occurred for all policies.
As a reference, we use the no-intervention policy $\govaction_0$ with no taxes, no integration subsidies, and the loosest license constraints. The reference values are the medians over eight equilibria of $\govaction_0$ computed from random initial strategies. 
\Cref{fig:pareto} shows the distribution of the municipality objective over the grid.
It ranges from $46{,}593$ to $48{,}951$~CHF/h, {and $89\%$ of the policies improve on $\govaction_0$ ($48{,}414$~CHF/h).
The best policy, with $\tax_\pt = -1$, $\tax_\taxi = 0.25$, $\subsidy_\pt = 9.2$~CHF/trip, $\subsidy_\taxi = 0$, and $\license = 147.5$~vehicles, achieves $46{,}593$~CHF/h}, while the median policy achieves $47{,}664$~CHF/h.

\begin{figure}[tb]
    \centering
    \includegraphics[width=0.85\linewidth]{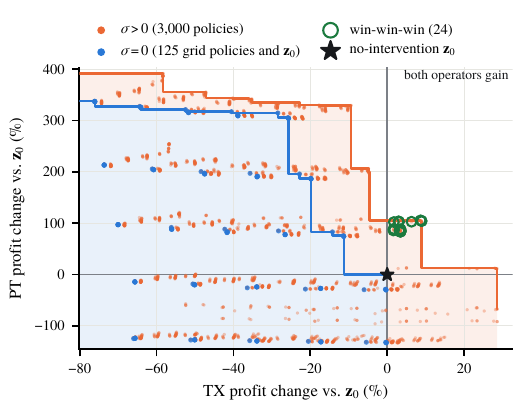}
    \caption{Percentage change in each operator's equilibrium profit relative to the no-intervention baseline ($\govaction_0$). The lines denote the Pareto frontiers for policies without integration subsidies (blue, comprising the $125$ grid policies and $\govaction_0$) and with integration subsidies (orange).
    Green circles mark the 24 globally Pareto-improving (GP) equilibria, labeled ``win-win-win'' in the legend, and the star marks~$\govaction_0$.}
    \label{fig:operator objective}
\end{figure}

\emph{Policies that benefit both operators:}
\Cref{fig:operator objective} shows each operator's profit at equilibrium relative to its value under $\govaction_0$ in percent, for policies without ($\subsidy = 0$) and with integration subsidies. 
With taxes and licenses alone, no policy improves both operators: the blue frontier reaches the origin only at $\govaction_0$ itself and never enters the upper-right quadrant. Integration subsidies extend the frontier into the upper-right quadrant, where both operators gain, by up to $29\%$ for \gls{acr:tx} and $106\%$ for \gls{acr:pt}.
In other words, taxes and licenses alone can improve one operator only at the expense of the other, whereas integration subsidies enlarge the set of outcomes that both operators prefer to~$\govaction_0$.

\emph{Analysis of globally Pareto-improving equilibria:}
We classify an equilibrium as \emph{operator Pareto-improving} (OP) if both operators improve upon the baseline $\govaction_0$, and \emph{globally Pareto-improving} (GP) if, in addition, travelers' costs decrease and the municipality objective does not increase. 
\Cref{tab:policy_comparison} counts these equilibria. 
Without integration subsidies, neither class occurs. 
With subsidies, $33$ policies are OP solutions and $24$ are GP solutions. 
All $24$ GP policies combine a low TX tax, a moderate PT revenue subsidy ($\tax_\pt = -0.25$), and a high TX integration subsidy ($\subsidy_\taxi \in \{9, 12\}$~CHF/trip).
Intuitively, the \gls{acr:tx} integration subsidy rewards \gls{acr:tx} for carrying riders who transfer from \gls{acr:pt}, so that \gls{acr:tx} complements transit rather than competing with it, which keeps its profit above the baseline.
Against $\govaction_0$, the median GP policy lowers the travel cost by $1{,}240$~CHF/h, raises the PT profit by $2{,}130$~CHF/h  and the TX profit by $190$~CHF/h, and improves the municipality objective by $590$~CHF/h.

\begin{table}[t]
\centering
\caption{Number of equilibria on the $5^5$ policy grid.}
\label{tab:policy_comparison}
\resizebox{\columnwidth}{!}{%
\begin{tabular}{lccc}
\toprule
Policy
& \makecell{All\\equilibria}
& \makecell{OP\\equilibria}
& \makecell{GP\\equilibria} \\
\midrule
No subsidy $\mathbf{\subsidy \!=\! 0}$  & 125     & 0 (0\%)      & 0 (0\%) \\
With subsidy $\mathbf{\subsidy \!>\! 0}$ & 3,000 & 33 (1.1\%)  & 24 (0.8\%) \\
\bottomrule
\end{tabular}
}
\end{table}

\subsubsection*{Top-Level Performance Optimization}
\label{sec:benchmark}
We compare the proposed controller, abbreviated as \gls{acr:mfo} against three black-box optimizers: \gls{acr:bo}~\cite{shahriari2016taking}, \gls{acr:CMA-ES}~\cite{hansen2003reducing}, and \gls{acr:ga}~\cite{katoch2021review}, all applied to the top-level control of the Zurich mobility network.
All methods initialize from $\govaction_0$, with the computational budget measured in evaluations of the municipal objective. 
The evaluation protocol comprises two stages using distinct random seeds: a tuning stage, where each configuration in \cref{tab:hyperparams} is evaluated for $200$ iterations across seeds $0$--$2$ to select the best-performing configuration (highlighted in bold in \cref{tab:hyperparams}); and a comparison stage, where the selected configuration is executed for $200$ iterations using 20 fresh seeds ($3$--$22$). 

\begin{table}[!tb]
\centering
\caption{Hyperparameter search ranges for different algorithms.}
\label{tab:hyperparams}
\begin{tabular}{llc}
\toprule
Algorithm & Hyperparameters & Ranges \\
\midrule
\multirow{2}{*}{BO}
  & Acquisition & EI, LCB $\kappa{=}2$, \textbf{LCB $\kappa$: 2$\to$0} \\
  & Initial points & $\{5,\; \mathbf{10},\; 25\}$ \\
\midrule
\multirow{2}{*}{CMA-ES}
  & Initial step size  & $\{\mathbf{0.1},\; 0.2,\; 0.5\}$ \\
  & Population size  & $\{\mathbf{8},\; 16,\; 32\}$ \\
\midrule
\multirow{2}{*}{GA}
  & Population size  & $\{10,\; \mathbf{20},\; 40\}$ \\
  & Mutation rate  & $\{0.1,\; \mathbf{0.2},\; 0.4\}$ \\
\midrule
\multirow{2}{*}{\gls{acr:mfo}}
  & Exploration radius $\delta_\govaction$ & $\{0.05,\; 0.1,\; 0.2, \;\mathbf{0.3}\}$ \\
  & Step size $\eta_\govaction$ & $\{1,\; 3,\;\mathbf{5},\; 10\}$ \\
\bottomrule
\end{tabular}
\begin{minipage}{\linewidth}
\footnotesize
\vspace{0.3em}
\textit{Note:} EI: expected improvement; LCB: lower confidence bound with weight $\kappa$, decreasing linearly over the run for ``2$\to$0''.
\end{minipage}
\end{table}

\begin{figure}[tb]
    \centering
    \includegraphics[width=0.95\linewidth]{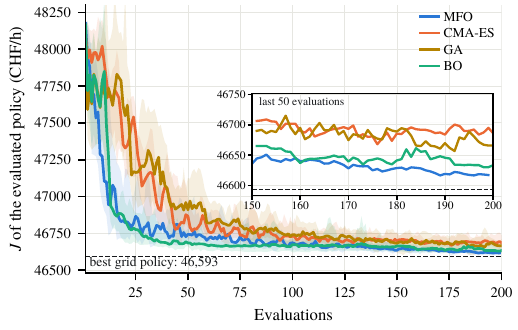}
    \caption{Objective of the evaluated policies over 200 evaluations. Lines show the mean over the 20 comparison runs, and bands the 25th to 75th percentiles.}
    \label{fig:benchmark}
\end{figure}

\begin{table}[t]
\centering
\caption{The last evaluated policy across 20 runs. Values indicate the median, best, worst, and standard deviation of the objective function, measured in CHF/h.}
\label{tab:benchmark_term}
\footnotesize
\begin{tabular}{lcccc}
\toprule
Method & Median & Best & Worst & Std. \\
\midrule
\textbf{MFO} & $\mathbf{46{,}614}$ & $46{,}591$ & $\mathbf{46{,}637}$ & $\mathbf{13}$ \\
CMA-ES       & $46{,}690$          & $46{,}594$ & $46{,}755$          & $43$          \\
GA           & $46{,}634$          & $46{,}571$ & $46{,}779$          & $61$          \\
BO           & $46{,}633$          & $46{,}609$ & $46{,}775$          & $39$          \\
\bottomrule
\end{tabular}
\end{table}

\Cref{fig:benchmark} shows the objective of the policies evaluated by each optimizer, and \cref{tab:benchmark_term} shows the objective of the terminal policy of each run, i.e., the last iterate for all methods.
All four methods start from $\govaction_0$. \gls{acr:bo} and \gls{acr:mfo} descend fastest, whereas \gls{acr:CMA-ES} and \gls{acr:ga} spend their first $60$ evaluations on widely spread policies (\cref{fig:benchmark}).
All four methods also reach the family of best grid policies. On average, the best objective found during a run lies between  $46{,}591$ and $46{,}609$~CHF/h for every method, a range well below the variation of the objective across initial operator strategies.
The best policies share the structure of the best grid policy, with full \gls{acr:pt} revenue support ($\tax_\pt \le -0.95$, median $-0.99$) and a moderate \gls{acr:tx} tax ($\tax_\taxi$ between $0.13$ and $0.24$).
They differ mainly in the \gls{acr:pt} integration subsidy ($\subsidy_\pt$ between $0.5$ and $9.2$~CHF/trip) and in the license ($106$ to $179$ vehicles), which suggests that the municipal objective is flat along these two directions near the optimum.

The methods differ in the policies at which they terminate.
\gls{acr:mfo} attains the lowest median terminal objective, $46{,}614$~CHF/h, against $46{,}690$~CHF/h for \gls{acr:CMA-ES} and $46{,}634$ and $46{,}633$~CHF/h for \gls{acr:ga} and \gls{acr:bo}, and the smallest standard deviation across runs, $13$~CHF/h, against $39$ to $61$~CHF/h. The worst run for \gls{acr:mfo} ($46{,}637$~CHF/h) is about the median of \gls{acr:bo} and \gls{acr:ga} (\cref{tab:benchmark_term}). 
We emphasize that the differences in median terminal objective, between $19$ and $76$~CHF/h, are small relative to the variation of the objective across operator initializations, so the four methods find policies of comparable quality.
The advantage of \gls{acr:mfo} lies instead in the predictability of its trajectory. Its iterates descend steadily toward the best policies and remain there, whereas \gls{acr:CMA-ES} and \gls{acr:ga} explore widely in the early evaluations and \gls{acr:bo} makes late-stage excursions.
This property matters when policies are adjusted incrementally on the real system rather than in a simulator, since every evaluated policy then reshapes the responses of operators and travelers.

Overall, the case study indicates that the operator loop computes reproducible equilibria, that integration subsidies are necessary to make both operators better off, and that model-free feedback attains the solution quality of established black-box optimizers while following a steady policy trajectory.

\section{Conclusion}\label{sec:conclusion}
In this article, we studied how a municipality can steer a multimodal mobility system whose operators and travelers respond strategically to its policies.
We modeled the system as a tri-level game, with a municipality above two competing operators and heterogeneous travelers in a stochastic user equilibrium, and we closed a feedback loop around the nested game.
We showed that the traveler equilibrium is unique and continuously differentiable in the operator strategies, that operator equilibria, characterized through a variational inequality, exist for every policy, and that these properties yield the existence of a Stackelberg equilibrium.
On this basis, we designed a hierarchical controller that relies only on zeroth-order evaluations of observed equilibrium outcomes, at both the municipal and the operator level.
On the Zurich network, the operator equilibria were reproducible across initializations, integration subsidies were necessary for policies that benefit both operators, and the controller reached the quality of the best policies found by black-box optimizers with the smallest run-to-run spread.

Several directions remain open.
First, our analysis establishes existence but not convergence. Deriving conditions under which the alternating zeroth-order operator updates converge to an operator equilibrium, and quantifying how inexact inner equilibria affect the municipal updates, would turn the empirical reliability observed in \cref{sec:numerical} into a guarantee. This requires controlling the monotonicity of the operator game, which the coupling through traveler flows can destroy.
Second, the operator game may admit multiple equilibria, and the controller implicitly selects one through its initialization. Characterizing this selection, or designing policies that perform well across all induced equilibria, matters whenever equilibria differ substantially in their societal cost.
Third, time-varying and stochastic demand, such as time-of-day patterns and day-to-day adaptation, would turn the plant into a dynamical system and bring the controller closer to online deployment, where steady policy trajectories become essential.
Fourth, the middle level can include further operators and modes, such as bike sharing, car sharing, or autonomous fleets, as well as cooperative structures such as revenue-sharing coalitions.
Finally, the municipality's problem can be posed as a multi-objective design problem that exposes the trade-offs between emissions, congestion, and social welfare, rather than fixing their relative weights a priori.

\appendices

\section{Additional Experimental Details}\label{sec:appendix}
\subsection{Model Parameters}
\color{black}
\Cref{table:model_parameters} summarizes the parameters across all three levels used in \cref{sec:numerical}. Values of time for commuting, business, and leisure travelers are calibrated from Swiss survey data. Municipal policy ranges define the feasible intervention set. We report detailed sources in~\cite[Table~I]{he2025hierarchical}.

\vspace{-0.7em}
\begin{table}[tbh]
    \centering
    \caption{Model parameters in the case study}
    \label{table:model_parameters}
    \resizebox{1\columnwidth}{!}{%
    \begin{tabular}{lllll}
        \hline
        \textbf{Parameter} & \textbf{Description} & \multicolumn{2}{c}{\textbf{Value}} & \textbf{Unit} \\ 
        \hline
        \multicolumn{5}{l}{\textbf{Travelers}} \\ 
        $\vot$ & Value of time [C, B, L] & \multicolumn{2}{c}{[19, 32, 12]} & CHF/h  \\
        $\theta$ & Logit dispersion & \multicolumn{2}{c}{10} & 1/h \\
        \multicolumn{2}{l}{\textbf{Operators}} & \textbf{TX} & \textbf{PT} & ~  \\
        $p^{\mathrm{base}}$ & Base price range & [0, 12] & [0, 9.2] & CHF \\
        $p^{\mathrm{d}}$ & Distance price range & [0, 8] & -- & CHF/km  \\
        $p^{\mathrm{t}}$ & Time price range & [0, 138] & -- & CHF/h \\
        $p^{\mathrm{trans}}$ & Transfer price range & [0, 12] & [0, 9.2] & CHF  \\
        $\ptfre$ & Service frequency range & -- & [2, 16] & veh/h \\
        $w$ & Fleet size range & [20, $\license$] & -- & veh \\
        $\costEnergy$ & Distance cost & 0.12 & 1.3 & CHF/km  \\
        $\costLabour$ & Time cost & 20 & 26 & CHF/h  \\
        $\costVehicle$ & Vehicle cost & 9 & 115 & CHF/veh/h  \\
        $\xi$ & Operator $\ell_2$ regularizer & 0 & 0 & -- \\
        $t^\mathrm{w}_0$ & Irreducible pickup time & 2 & -- & min \\
        $t^\mathrm{w}_\text{m}$ & Matching wait at $\bar w$ & 4 & -- & min \\
        $\bar w$ & Reference fleet size & 500 & -- & veh \\
        \multicolumn{2}{l}
        {\textbf{Municipality}} & \textbf{TX} & \textbf{PT} & \\
        $\tax$ & Revenue tax range & [0, 0.5] & [$-$1, 0.5] & --  \\
        $\subsidy$ & Integration subsidy range & [0, 12] & [0, 9.2] & CHF/trip \\
        $\license$ & Fleet license range & [50, 180] & -- & veh \\
        $\unitEmission$ & CO$_2$ emission rate & \multicolumn{2}{c}{0.15} & kg/km  \\
        $\omega_1, \omega_2, \omega_3$ & Weights in~\eqref{eq:municipality_prob} & \multicolumn{2}{c}{1, 0.12, 0.2} & --, CHF/kg, --\\
        \hline
    \end{tabular}
    }
\end{table}

\subsection{Convergence of the Operator Loop}\label{app:www}
\begin{figure*}[tb]
    \centering
    \includegraphics[width=0.85\textwidth]{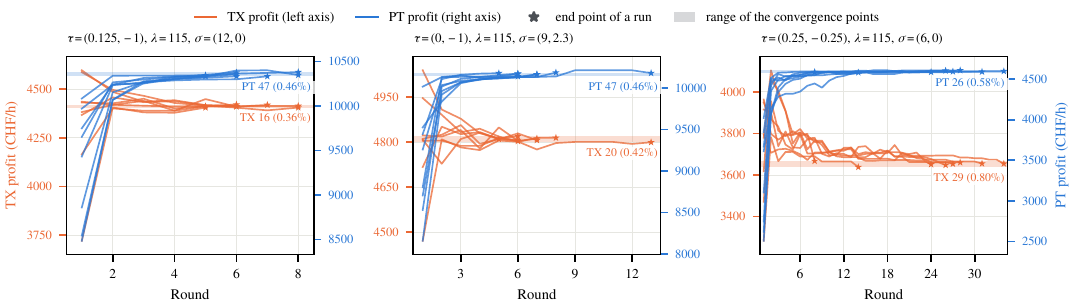}
    \caption{Operator profits along the rounds of the operator loop for three sampled policies with $\varepsilon_\mathrm{MNE} = 0.03$, eight random initial strategies each.}
    \label{fig:br_unique}
\end{figure*}

\Cref{fig:br_unique} illustrates the convergence behavior for three of the policies sampled in \cref{sec:reliability}. Despite the profits of initial strategies differing by as much as $900$~CHF/h (TX) and $1{,}800$~CHF/h (PT), the trajectories consistently lead to the same equilibrium. Across the eight initializations, the final profit values lie within $0.8\%$ of the median for TX and $0.6\%$ for PT. 

\makeatletter\def\@textbottom{\vskip\z@\@plus 1fill}\let\@texttop\relax\makeatother
\makeatletter\def\@IEEEBIOskipN{1\baselineskip}\makeatother
\makeatletter\def\@IEEEBIOskipN{1\baselineskip}\def\@IEEEBIOhangdepth{1in}\def\@IEEEBIOphotodepth{1in}\makeatother

\bibliographystyle{IEEEtran}
\bibliography{reference}

\begin{thebibliography}{10}
\providecommand{\url}[1]{#1}
\csname url@samestyle\endcsname
\providecommand{\newblock}{\relax}
\providecommand{\bibinfo}[2]{#2}
\providecommand{\BIBentrySTDinterwordspacing}{\spaceskip=0pt\relax}
\providecommand{\BIBentryALTinterwordstretchfactor}{4}
\providecommand{\BIBentryALTinterwordspacing}{\spaceskip=\fontdimen2\font plus
\BIBentryALTinterwordstretchfactor\fontdimen3\font minus \fontdimen4\font\relax}
\providecommand{\BIBforeignlanguage}[2]{{%
\expandafter\ifx\csname l@#1\endcsname\relax
\typeout{** WARNING: IEEEtran.bst: No hyphenation pattern has been}%
\typeout{** loaded for the language `#1'. Using the pattern for}%
\typeout{** the default language instead.}%
\else
\language=\csname l@#1\endcsname
\fi
#2}}
\providecommand{\BIBdecl}{\relax}
\BIBdecl

\bibitem{eu2020mobilitystrategy}
\BIBentryALTinterwordspacing
{European Commission}, ``Sustainable and smart mobility strategy,'' European Commission, Brussels, Tech. Rep. COM(2020) 789 final, 2020. [Online]. Available: \url{https://eur-lex.europa.eu/legal-content/EN/TXT/?uri=CELEX:52020DC0789}
\BIBentrySTDinterwordspacing

\bibitem{cianfanelli2023optimal}
L.~Cianfanelli, G.~Como, A.~E. Ozdaglar, and F.~Parise, ``Optimal intervention in transportation networks,'' \emph{IEEE Transactions on Automatic Control}, vol.~68, no.~12, pp. 7073--7088, 2023.

\bibitem{grontas2023designing}
P.~D. Grontas, C.~Cenedese, M.~Fochesato, G.~Belgioioso, J.~Lygeros, and F.~D{\"o}rfler, ``Designing optimal personalized incentive for traffic routing using big hype,'' in \emph{2023 62nd IEEE Conference on Decision and Control (CDC)}.\hskip 1em plus 0.5em minus 0.4em\relax IEEE, 2023, pp. 3142--3147.

\bibitem{mehr2019will}
N.~Mehr and R.~Horowitz, ``How will the presence of autonomous vehicles affect the equilibrium state of traffic networks?'' \emph{IEEE Transactions on Control of Network Systems}, vol.~7, no.~1, pp. 96--105, 2019.

\bibitem{kim2025strategic}
Y.~Kim, N.~Duan, G.~Zardini, S.~Samaranayake, and D.~Wischik, ``Strategic pricing and routing to maximize profit in congested roads considering interactions with travelers,'' \emph{IEEE Transactions on Control of Network Systems}, 2025.

\bibitem{zardini2021game}
G.~Zardini, N.~Lanzetti, L.~Guerrini, E.~Frazzoli, and F.~D{\"o}rfler, ``Game theory to study interactions between mobility stakeholders,'' in \emph{2021 IEEE International Intelligent Transportation Systems Conference (ITSC)}.\hskip 1em plus 0.5em minus 0.4em\relax IEEE, 2021, pp. 2054--2061.

\bibitem{zardini2023strategic}
G.~Zardini, N.~Lanzetti, G.~Belgioioso, C.~Hartnik, S.~Bolognani, F.~D{\"o}rfler, and E.~Frazzoli, ``Strategic interactions in multi-modal mobility systems: A game-theoretic perspective,'' in \emph{2023 IEEE 26th International Conference on Intelligent Transportation Systems (ITSC)}.\hskip 1em plus 0.5em minus 0.4em\relax IEEE, 2023, pp. 5452--5459.

\bibitem{lanzetti2023interplay}
N.~Lanzetti, M.~Schiffer, M.~Ostrovsky, and M.~Pavone, ``On the interplay between self-driving cars and public transportation,'' \emph{IEEE Transactions on Control of Network Systems}, vol.~11, no.~3, pp. 1478--1490, 2023.

\bibitem{maljkovic2025decentralized}
M.~Maljkovic, G.~Nilsson, and N.~Geroliminis, ``On decentralized computation of the leader’s strategy in bi-level games,'' \emph{Automatica}, vol. 178, p. 112352, 2025.

\bibitem{li2026implementation}
X.~Li, R.~Zhang, and G.~Zardini, ``Implementation-based incentive design for autonomous mobility-on-demand and transit systems,'' \emph{arXiv preprint arXiv:2605.18687}, 2026.

\bibitem{zambrano2025user}
M.~Zambrano, X.~Li, R.~Fiorista, and G.~Zardini, ``User-friendly game-theoretic modeling and analysis of multi-modal transportation systems,'' in \emph{2025 IEEE Conference on Control Technology and Applications (CCTA)}.\hskip 1em plus 0.5em minus 0.4em\relax IEEE, 2025, pp. 723--728.

\bibitem{zhang2023ride}
K.~Zhang, A.~Mittal, S.~Djavadian, R.~Twumasi-Boakye, and Y.~M. Nie, ``Ride-hail vehicle routing (river) as a congestion game,'' \emph{Transportation Research Part B: Methodological}, vol. 177, p. 102819, 2023.

\bibitem{sheldon2025competitive}
M.~Sheldon, D.~Paccagnan, and G.~Casale, ``Competitive analysis of vehicle-sharing systems with cournot queueing games,'' \emph{IEEE Transactions on Intelligent Transportation Systems}, 2025.

\bibitem{he2025co}
M.~He, A.~Censi, E.~Frazzoli, and G.~Zardini, ``Co-investment with payoff sharing benefit operators and users in network design,'' in \emph{2025 American Control Conference (ACC)}.\hskip 1em plus 0.5em minus 0.4em\relax IEEE, 2025, pp. 1131--1138.

\bibitem{mo2021competition}
B.~Mo, Z.~Cao, H.~Zhang, Y.~Shen, and J.~Zhao, ``Competition between shared autonomous vehicles and public transit: A case study in singapore,'' \emph{Transportation Research Part C: Emerging Technologies}, vol. 127, p. 103058, 2021.

\bibitem{maljkovic2023hierarchical}
M.~Maljkovic, G.~Nilsson, and N.~Geroliminis, ``Hierarchical pricing game for balancing the charging of ride-hailing electric fleets,'' \emph{IEEE Transactions on Control Systems Technology}, vol.~31, no.~6, pp. 2728--2743, 2023.

\bibitem{gu2019tri}
Y.~Gu, X.~Cai, D.~Han, and D.~Z. Wang, ``A tri-level optimization model for a private road competition problem with traffic equilibrium constraints,'' \emph{European Journal of Operational Research}, vol. 273, no.~1, pp. 190--197, 2019.

\bibitem{xi2024single}
H.~Xi, D.~Aussel, W.~Liu, S.~T. Waller, and D.~Rey, ``Single-leader multi-follower games for the regulation of two-sided mobility-as-a-service markets,'' \emph{European Journal of Operational Research}, vol. 317, no.~3, pp. 718--736, 2024.

\bibitem{dandl2021regulating}
F.~Dandl, R.~Engelhardt, M.~Hyland, G.~Tilg, K.~Bogenberger, and H.~S. Mahmassani, ``Regulating mobility-on-demand services: Tri-level model and {B}ayesian optimization solution approach,'' \emph{Transportation Research Part C: Emerging Technologies}, vol. 125, p. 103075, 2021.

\bibitem{yao2026maas}
R.~Yao, X.~Ma, and K.~Zhang, ``Mobility-as-a-service (maas) system as a multi-leader-multi-follower game: A single-level variational inequality (vi) formulation,'' \emph{Transportation Research Part C: Emerging Technologies}, vol. 188, p. 105703, 2026.

\bibitem{bacsar1998dynamic}
T.~Ba{\c{s}}ar and G.~J. Olsder, \emph{Dynamic noncooperative game theory}.\hskip 1em plus 0.5em minus 0.4em\relax SIAM, 1998.

\bibitem{grontas2024big}
P.~D. Grontas, G.~Belgioioso, C.~Cenedese, M.~Fochesato, J.~Lygeros, and F.~D{\"o}rfler, ``Big hype: Best intervention in games via distributed hypergradient descent,'' \emph{IEEE Transactions on Automatic Control}, vol.~69, no.~12, pp. 8338--8353, 2024.

\bibitem{belgioioso2020distributed}
G.~Belgioioso, A.~Nedi{\'c}, and S.~Grammatico, ``Distributed generalized nash equilibrium seeking in aggregative games on time-varying networks,'' \emph{IEEE Transactions on Automatic Control}, vol.~66, no.~5, pp. 2061--2075, 2020.

\bibitem{yi2019operator}
P.~Yi and L.~Pavel, ``An operator splitting approach for distributed generalized nash equilibria computation,'' \emph{Automatica}, vol. 102, pp. 111--121, 2019.

\bibitem{simonetto2020time}
A.~Simonetto, E.~Dall'Anese, S.~Paternain, G.~Leus, and G.~B. Giannakis, ``Time-varying convex optimization: Time-structured algorithms and applications,'' \emph{Proceedings of the IEEE}, vol. 108, no.~11, pp. 2032--2048, 2020.

\bibitem{hauswirth2024optimization}
A.~Hauswirth, Z.~He, S.~Bolognani, G.~Hug, and F.~D{\"o}rfler, ``Optimization algorithms as robust feedback controllers,'' \emph{Annual Reviews in Control}, vol.~57, 2024, {A}rt.~no.~100941.

\bibitem{belgioioso2025online}
G.~Belgioioso, D.~Liao-McPherson, M.~H. de~Badyn, S.~Bolognani, R.~S. Smith, J.~Lygeros, and F.~D{\"o}rfler, ``Online feedback equilibrium seeking,'' \emph{IEEE Transactions on Automatic Control}, vol.~70, no.~1, pp. 203--218, 2025.

\bibitem{bai2021sample}
Y.~Bai, C.~Jin, H.~Wang, and C.~Xiong, ``Sample-efficient learning of stackelberg equilibria in general-sum games,'' \emph{Advances in Neural Information Processing Systems}, vol.~34, pp. 25\,799--25\,811, 2021.

\bibitem{kwon2023fully}
J.~Kwon, D.~Kwon, S.~Wright, and R.~D. Nowak, ``A fully first-order method for stochastic bilevel optimization,'' in \emph{International Conference on Machine Learning}.\hskip 1em plus 0.5em minus 0.4em\relax PMLR, 2023, pp. 18\,083--18\,113.

\bibitem{he2025hierarchical}
M.~He, Z.~He, J.~Ghadamian, F.~D{\"o}rfler, E.~Frazzoli, and G.~Zardini, ``Hierarchical strategic decision-making in layered mobility systems,'' in \emph{European Control Conference}, 2026.

\bibitem{BPR}
N.~Gore, S.~Arkatkar, G.~Joshi, and C.~Antoniou, ``Modified bureau of public roads link function,'' \emph{Transportation Research Record}, vol. 2677, no.~5, pp. 966--990, 2023.

\bibitem{ingvardson2018passenger}
J.~B. Ingvardson, O.~A. Nielsen, S.~Raveau, and B.~F. Nielsen, ``Passenger arrival and waiting time distributions dependent on train service frequency and station characteristics: A smart card data analysis,'' \emph{Transportation Research Part C: Emerging Technologies}, vol.~90, pp. 292--306, 2018.

\bibitem{castillo2025matching}
J.~C. Castillo, D.~Knoepfle, and E.~G. Weyl, ``Matching and pricing in ride hailing: Wild goose chases and how to solve them,'' \emph{Management Science}, vol.~71, no.~5, pp. 4377--4395, 2025.

\bibitem{Axhausen2006SwissVTT}
K.~W. Axhausen, S.~Hess, A.~König, G.~Abay, J.~Bates, and M.~Bierlaire, ``State of the art estimates of the swiss value of travel time savings,'' vol. 383.\hskip 1em plus 0.5em minus 0.4em\relax ETH Zürich, 2006.

\bibitem{Boyles2022}
S.~Boyles, N.~E. Lownes, and A.~Unnikrishnan, \emph{Transportation Network Analysis}, 2022, vol.~1.

\bibitem{dontchev2009implicit}
A.~L. Dontchev and R.~T. Rockafellar, \emph{Implicit functions and solution mappings}.\hskip 1em plus 0.5em minus 0.4em\relax New York, NY, USA: Springer, 2009, vol. 543.

\bibitem{fiacco1990sensitivity}
A.~V. Fiacco and Y.~Ishizuka, ``Sensitivity and stability analysis for nonlinear programming,'' \emph{Annals of Operations Research}, vol.~27, no.~1, pp. 215--235, 1990.

\bibitem{mertikopoulos2019learning}
P.~Mertikopoulos and Z.~Zhou, ``Learning in games with continuous action sets and unknown payoff functions,'' \emph{Mathematical Programming}, vol. 173, no.~1, pp. 465--507, 2019.

\bibitem{facchinei2003finite}
F.~Facchinei and J.-S. Pang, \emph{Finite-dimensional variational inequalities and complementarity problems}.\hskip 1em plus 0.5em minus 0.4em\relax Springer, 2003.

\bibitem{aliprantis2006infinite}
C.~D. Aliprantis and K.~C. Border, \emph{Infinite dimensional analysis: a hitchhiker’s guide}.\hskip 1em plus 0.5em minus 0.4em\relax Springer, 2006.

\bibitem{blondel2024elements}
M.~Blondel and V.~Roulet, ``The elements of differentiable programming,'' \emph{arXiv preprint arXiv:2403.14606}, 2024.

\bibitem{Zhiyu2024}
Z.~He, S.~Bolognani, J.~He, F.~Dörfler, and X.~Guan, ``Model-free nonlinear feedback optimization,'' \emph{IEEE Transactions on Automatic Control}, vol.~69, no.~7, pp. 4554--4569, 2024.

\bibitem{liu2009method}
H.~X. Liu, X.~He, and B.~He, ``Method of successive weighted averages (mswa) and self-regulated averaging schemes for solving stochastic user equilibrium problem,'' \emph{Networks and Spatial Economics}, vol.~9, no.~4, pp. 485--503, 2009.

\bibitem{openstreetmap}
{OpenStreetMap}, \url{https://www.openstreetmap.org}, 2025.

\bibitem{sbfsPortal}
\BIBentryALTinterwordspacing
{Swiss Federal Statistical Office}. (2025) Statistics. [Online]. Available: \url{https://www.bfs.admin.ch/bfs/en/home/statistics.html}
\BIBentrySTDinterwordspacing

\bibitem{loh1996latin}
W.-L. Loh, ``On latin hypercube sampling,'' \emph{The annals of statistics}, vol.~24, no.~5, pp. 2058--2080, 1996.

\bibitem{shahriari2016taking}
B.~Shahriari, K.~Swersky, Z.~Wang, R.~P. Adams, and N.~de~Freitas, ``Taking the human out of the loop: {A} review of {B}ayesian optimization,'' \emph{Proceedings of the IEEE}, vol. 104, no.~1, pp. 148--175, 2016.

\bibitem{hansen2003reducing}
N.~Hansen and A.~Ostermeier, ``Completely derandomized self-adaptation in evolution strategies,'' \emph{Evolutionary Computation}, vol.~9, no.~2, pp. 159--195, 2001.

\bibitem{katoch2021review}
S.~Katoch, S.~S. Chauhan, and V.~Kumar, ``A review on genetic algorithm: past, present, and future,'' \emph{Multimedia Tools and Applications}, vol.~80, no.~5, pp. 8091--8126, 2021.

\end{thebibliography}

{\linespread{0.95}\begin{IEEEbiography}[{\includegraphics[width=1in,height=1in,clip,keepaspectratio]{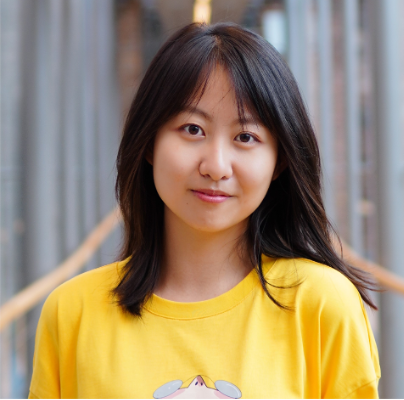}}]{Mingjia He}
is a doctoral student at the Institute for Dynamic Systems and Control, ETH Zürich. 
Her research focuses on strategic interactions in multi-agent autonomous systems, aiming to develop control strategies and cooperative mechanisms for future mobility. 
She received the M.Sc. degree from Delft University of Technology and the B.Sc. degree from Southeast University.
\end{IEEEbiography}

\begin{IEEEbiography}
[{\includegraphics[width=1in,height=1in,clip,keepaspectratio]{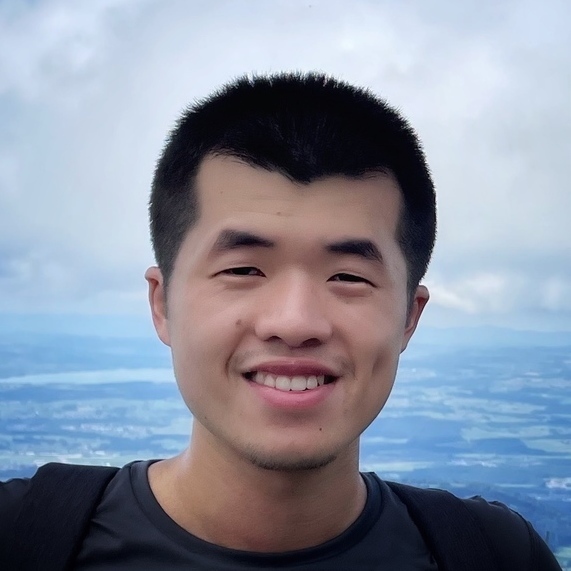}}]{Zhiyu He} 
received the B.Eng.\ degree and the M.S.\ degree in Control Science and Engineering from Shanghai Jiao Tong University, Shanghai, China, in 2019 and 2022, respectively. He is currently a Ph.D.\ student at the Automatic Control Laboratory at ETH Z{\"u}rich. His research interests lie in optimization, learning, and control of dynamical systems.
\end{IEEEbiography}

\vspace{-1ex}

\begin{IEEEbiography}[{\includegraphics[width=1in,height=1in,clip,keepaspectratio]{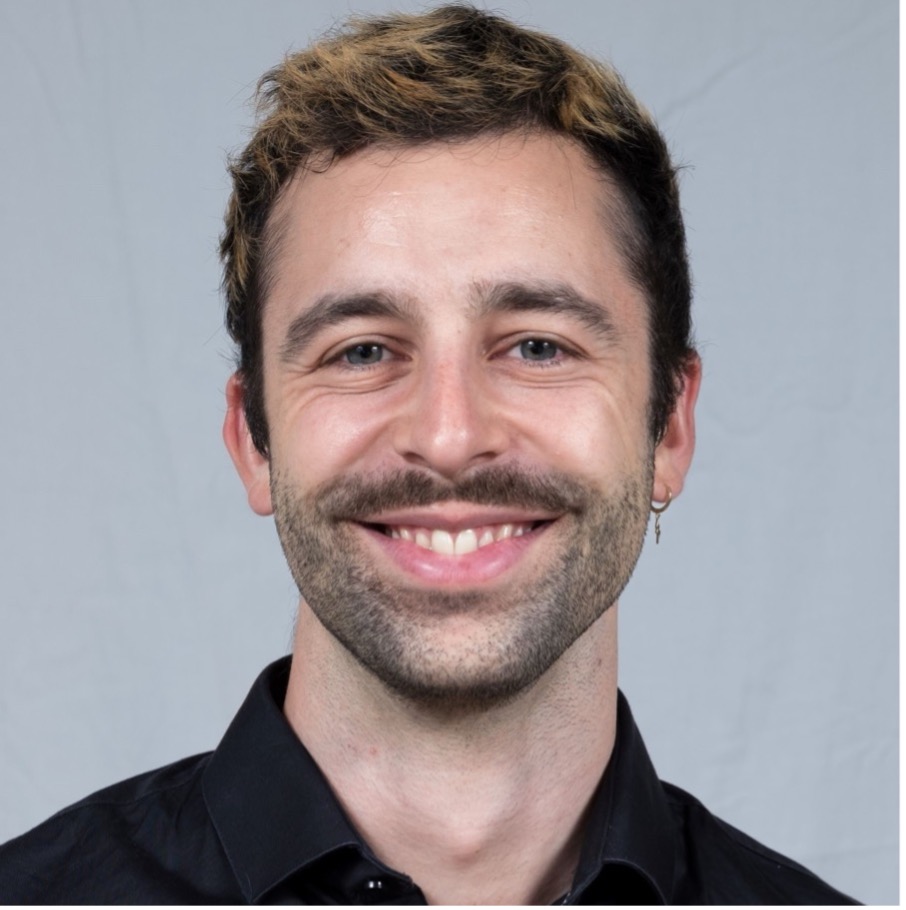}}]{Jan Ghadamian}
is a doctoral student at the Institute of Construction and Infrastructure Management, ETH Z{\"u}rich, under the supervision of Prof. Bryan T. Adey.
His research focuses on infrastructure systems modeling, transportation network analysis, and decision-making under uncertainty for urban mobility systems.
\end{IEEEbiography}

\begin{IEEEbiography}[{\includegraphics[width=1in,height=1in,clip,keepaspectratio]{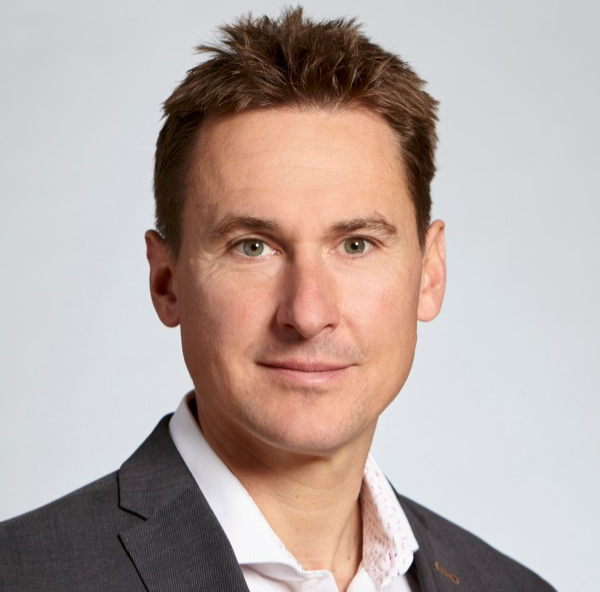}}]{Florian D\"{o}rfler}
received the Diploma in engineering cybernetics from the University of Stuttgart, Stuttgart, Germany, in 2008 and the Ph.D. degree in mechanical engineering from the University of California at Santa Barbara, CA, USA, in 2013.
He is a Full Professor with the Automatic Control Laboratory, ETH Z{\"u}rich, Z{\"u}rich, Switzerland. From 2013 to 2014, he was an Assistant Professor with the University of California, Los Angeles, CA, USA. He served as the Associate Head of the Department of Information Technology and Electrical Engineering, ETH Z{\"u}rich, from 2021 until 2022. 
His research interests include control, optimization, and system theory with applications in network systems, in particular electric power grids.
\end{IEEEbiography}

\begin{IEEEbiography}[{\includegraphics[width=1in,height=1in,clip,keepaspectratio]{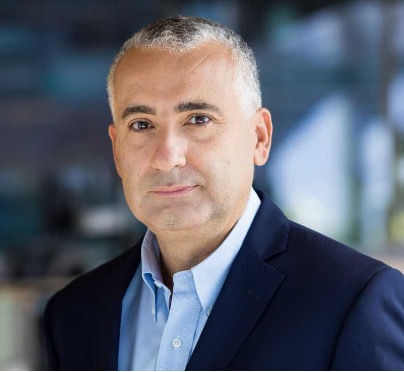}}]{Emilio Frazzoli} 
is a Professor of Dynamic Systems and Control at ETH Z\"{u}rich.
Until March 2021, he was Chief Scientist of Motional, the latest embodiment of nuTonomy, the startup he co-founded in 2013. His main research interests are in robotics, autonomous systems, and intelligent mobility. In acknowledgment of his work in these fields, he has received several awards, including the 2015 IEEE George S. Axelby Award, the 2017 IEEE Kiyo Tomiyasu Award, the 2022 IEEE George N. Saridis Award, and has been named an IEEE Fellow in 2019.  He holds a Laurea Degree in Aeronautical Engineering from the Sapienza University of Rome, and a PhD in Aeronautics and Astronautics from MIT.\end{IEEEbiography}

\begin{IEEEbiography}[{\includegraphics[width=1in,height=1in,clip,keepaspectratio]{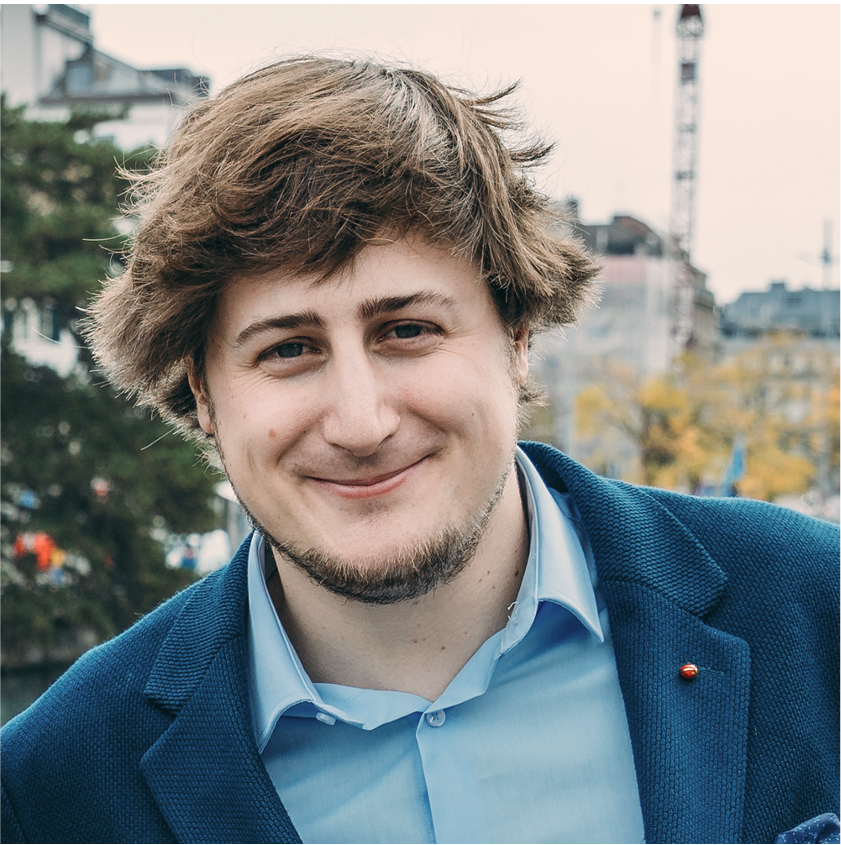}}]{Gioele Zardini} 
is the Rudge (1948) and Nancy Allen Assistant Professor at MIT, where he is a PI in LIDS and CEE and an affiliate faculty in IDSS. He received the Ph.D. degree from ETH Z{\"u}rich in 2024. His interests include the co-design of complex systems, all the way from future mobility to autonomous systems, compositionality, and control theory architectures. 
He leads the AFOSR Center of Excellence on Compositional Optimization, Dynamical Systems, and Control. 
He was awarded the ETH Medal for his doctoral thesis, the 2025 DARPA Young Faculty Award, and the 2026 Maseeh Excellence in Teaching Award at MIT.
\end{IEEEbiography}
\vfill
}

\end{document}